\documentclass[journal]{IEEEtran}
\usepackage{amssymb}
\usepackage[cmex10]{amsmath}
\usepackage{stfloats}
\usepackage{graphicx}
\usepackage{subfigure}
\usepackage{tabularx}
\usepackage{epsfig,epsf,color,balance,cite}
\usepackage{verbatim}
\usepackage{url}
\usepackage{bm}
\usepackage{mathrsfs}
\usepackage{amsthm}
\DeclareMathAlphabet\mathbfcal{OMS}{cmsy}{b}{n}

\newtheorem{theorem}{Theorem}

\newtheorem{lemma}{Lemma}
\newtheorem{remark}{Remark}
\usepackage{algorithm}
\usepackage{algpseudocode}
\usepackage{graphicx}
\usepackage{epstopdf}
\IEEEoverridecommandlockouts

\begin{document}

\title{Achievable Regions and Precoder Designs for the Multiple Access Wiretap Channels with Confidential and Open Messages}

\author{
\IEEEauthorblockN{Hao Xu, \emph{Member, IEEE}\IEEEauthorrefmark{0}, 
	              Tianyu Yang, \IEEEauthorrefmark{0}     
	              Kai-Kit Wong, \emph{Fellow, IEEE}\IEEEauthorrefmark{0},
and
	              Giuseppe Caire, \emph{Fellow, IEEE}\IEEEauthorrefmark{0}
}
 \thanks{
	H. Xu and G. Caire were supported by the Alexander von Humboldt Foundation.
	Partial of this work has been presented in 2020 IEEE International Symposium on Information Theory (ISIT) \cite{9174164}. (Corresponding author: Hao Xu)
	
	H. Xu, T. Yang and, G. Caire are with the Faculty of Electrical Engineering and Computer Science at the Technical University of Berlin, 10587 Berlin, Germany (e-mail: xuhao@mail.tu-berlin.de; tianyu.yang@tu-berlin.de; caire@tu-berlin.de).
	
	K.-K. Wong is with the Department of Electronic and Electrical Engineering, University College London, London WC1E7JE, UK (e-mail: kai-kit.wong@ucl.ac.uk).
}
}

\maketitle

\begin{abstract}
This paper investigates the secrecy capacity region of multiple access wiretap (MAC-WT) channels where, besides confidential messages, the users have also open messages to transmit.
All these messages are intended for the legitimate receiver (or Bob for brevity) but only the confidential messages need to be protected from the eavesdropper (Eve). 
We first consider a discrete memoryless (DM) MAC-WT channel where both Bob and Eve jointly decode their interested messages.
By using random coding, we find an achievable rate region, within which perfect secrecy can be realized, i.e., all users can communicate with Bob with arbitrarily small probability of error, while the confidential information leaked to Eve tends to zero.
Due to the high implementation complexity of joint decoding, we also consider the DM MAC-WT channel where Bob simply decodes messages independently while Eve still applies joint decoding.
We then extend the results in the DM case to a Gaussian vector (GV) MAC-WT channel.
Based on the information theoretic results, we further maximize the sum secrecy rate of the GV MAC-WT system by designing precoders for all users.
Since the problems are non-convex, we provide iterative algorithms to obtain suboptimal solutions.
Simulation results show that compared with existing schemes, secure communication can be greatly enhanced by the proposed algorithms, and in contrast to the works which only focus on the network secrecy performance, the system spectrum efficiency can be effectively improved since open messages can be simultaneously transmitted.

\end{abstract}


\IEEEpeerreviewmaketitle

\section{Introduction}
\label{section1}

With the rapid development of mobile communication, more and more sensitive and confidential information is sent over the air. 
Due to its broadcast nature, wireless transmission is inherently vulnerable to security breaches, which has been a pivotal issue in the modern wireless communication systems \cite{yang2015safeguarding, 7762075}.
It is thus of great importance to enhance the wireless transmission security. 
Traditionally, secure wireless communication is guaranteed by key-based cryptographic techniques, which rely on secret keys and assumptions of limited computational ability of the eavesdroppers (Eves).
Since the future mobile network will incorporate different network topologies and large numbers of mobile devices which may access and leave at any time, it will be difficult to generate and manage cryptographic keys. 
In addition, owing to the unprecedented growth of computational ability, the Eves may be able to extract the confidential information of authorized users without secret keys.
Hence, the conventional cryptographic encryption methods are no longer sufficient to guarantee secure communications in the future mobile networks.

To make up for these shortcomings, advanced signal processing techniques developed for facilitating security in the physical layer have emerged and triggered considerable research interests in recent years \cite{7762075}.
Different from the conventional cryptographic encryption methods employed in the application layer, physical layer security techniques exploit the random propagation properties of radio channels to prevent Eves from wiretapping.
The study of information theoretic secrecy in communications starts from some seminal works \cite{shannon1949communication, wyner1975wire, leung1978gaussian, csiszar1978broadcast}.
Specifically, in \cite{shannon1949communication}, Shannon first gave a rigorous analysis of information theoretic secrecy, based on which Wyner defined the wiretap channel in \cite{wyner1975wire}.
In work \cite{wyner1975wire}, Wyner considered a discrete memoryless (DM) channel with an Eve which has access to a stochastically degraded version of the main channel's output, and aimed to maximize the transmission rate to the legitimate receiver while keeping the Eve as ignorant of the secret message as possible.
Based on \cite{wyner1975wire}, the achievable rate-equivocation region of a degraded Gaussian wiretap channel was investigated in \cite{leung1978gaussian}. 
In \cite{csiszar1978broadcast}, Wyner's work was extended to a non-degraded wiretap channel, and to a scenario including a confidential message for the legitimate receiver only and a common message intended for both the legitimate receiver and the Eve.

Ever since these pioneering works \cite{shannon1949communication, wyner1975wire, leung1978gaussian, csiszar1978broadcast}, the research of wiretap channels or physical layer security has evolved into various network topologies over the past decades, e.g., multiple access (MAC) wiretap channels \cite{ekrem2008secrecy, tekin2008gaussian, tekin2008general, 4036106, 4455769, 5961828, nafea2019generalizing}, broadcast channels \cite{csiszar1978broadcast, 5730586, 5550390, 5605348, 6584931, 9133130}, interference channels \cite{4529283, 4595013, 5752448, 6006610, 7060726, 7313047}, relay-aided channels \cite{955145, 4608977, 5352243, 6601774, 7105936, 7355564, 7551149}, etc.
As one of the earliest and most important channels, MAC channels have been widely studied in the literatures \cite{Ahlswede1971multi, Liao1972Multiple}.
In the next generation network, to accomplish heterogeneous services and applications, e.g., virtual reality (VR), augmented reality (AR), holographic telepresence, industry 4.0, etc., it is of great importance to study MAC problems and enhance the performance of MAC systems in terms of massive user access, high bandwidth efficiency, low latency services, and transmission secrecy.
This paper mainly studies the physical layer security problem in a MAC wiretap (MAC-WT) channel.
Hence, we introduce MAC-WT related works in the following.

Both \cite{ekrem2008secrecy} and \cite{tekin2008gaussian} considered a MAC-WT channel with a weaker eavesdropper which has access to a degraded version of the main channel.  
Specifically, reference \cite{ekrem2008secrecy} considered a DM MAC-WT channel and developed an outer bound for the secrecy capacity region.
Reference \cite{tekin2008gaussian} first defined two separate secrecy measures for a Gaussian MAC-WT channel and then provided achievable rate regions under different secrecy constraints.
The work in \cite{tekin2008gaussian} was extended to a more general non-degraded MAC-WT channel by \cite{tekin2008general}.
In \cite{tekin2008general}, each user had a secret and an open message to transmit, and an achievable rate region for both secret and open rates was provided. 
In \cite{4036106}, the authors consider a two-user DM MAC channel in which one user wishes to communicate confidential messages to a common receiver while the other user is permitted to eavesdrop.
The upper bounds and achievable rates for this communication situation were determined.
A similar two-user DM MAC system was studied in \cite{4455769}.
Differently, each user in \cite{4455769} attempts to transmit both common and confidential information to the destination, and views the other user as an Eve. 
Inner and outer bounds on the capacity-equivocation region and secrecy capacity region were investigated in \cite{4455769}.
Reference \cite{5961828} extended the work in \cite{4036106} and \cite{4455769} to a fading cognitive MAC channel.
In \cite{nafea2019generalizing}, the MAC-WT channel with a DM main channel and different wiretapping scenarios were studied.

Based on these information theoretic secrecy results, a lot of work has studied the resource allocation problems in MAC-WT channels \cite{tekin2008general, hao2018resource, 8895802, lee2017precoder}.
Using the derived achievable rate region, the sum secrecy rate of a single-input single-output (SISO) Gaussian MAC-WT channel was maximized by power control in \cite{tekin2008general}.
It was shown in \cite[Theorem 3]{tekin2008general} that in the optimal case, only a subset of the strong users will transmit using the maximum power while the other users are inactive.
In \cite{hao2018resource} and \cite{8895802}, the sum secrecy rate maximization problems of a single-input multi-output (SIMO) heterogeneous network, i.e., a network with both conventional mobile users in uplink mode and device-to-device (D2D) pairs, were studied.
Obviously, the heterogeneous network can be seen as a MAC-WT channel when there is no D2D users.
In \cite{hao2018resource}, D2D users acted as friendly jammers and the sum secrecy rate of all mobile users was maximized, while in \cite{8895802}, the secrecy performance of all users was considered by maximizing the weighted sum secrecy rate.
However, it was assumed in \cite{hao2018resource} and \cite{8895802} that both the base station, which is the legitimate receiver, and the Eve adopt linear detection, i.e., treating the interference from the other users as independent additive noise when decoding the signal of one user.
The MAC-WT system thus reduces to a set of point-to-point wiretap channels.
However, designing the system under the assumption that the Eve is constrained to using a suboptimal detection scheme may be risky, since the system secrecy may break down if the Eve makes use of an enhanced receiver. 
In \cite{lee2017precoder}, the sum secrecy rate maximization problem of a multi-input multi-output (MIMO) MAC-WT system was studied and it was assumed that both the legitimate receiver and the Eve apply the optimal decoding scheme, i.e., joint decoding.
However, the authors set a special power constraint to the covariance matrices of the transmit signal vectors in \cite{lee2017precoder}, which may limit the secrecy performance of the network.

In this paper, we study the information theoretic secrecy problem for a general MAC-WT channel.
Different from \cite{tekin2008general}, which studied secrecy communication of a SISO Gaussian MAC-WT channel, we first derive achievable rate regions for a DM MAC-WT channel and then extend the results to the Gaussian vector (GV) MAC-WT channel.
Besides the confidential message, each user also has an open message intended for the legitimate receiver (or Bob for brevity).
This constitutes a generalization of the results in \cite{nafea2019generalizing}, where each user only transmits a secret message, and helps improve the network spectrum efficiency since in contrast to the works which only focus on the network secrecy performance, more (open) information can be transmitted.
The main contributions of this paper are summarized as follows:

$\bullet$ We first consider a DM MAC-WT channel and assume that both Bob and Eve apply the joint decoding method to detect their interested messages.
By using random coding, we find an achievable rate region for the channel, where users can communicate with Bob with arbitrarily small probability of error, while the confidential information leaked to Eve tends to zero.
Since the complexity of joint decoding grows exponentially with the signal dimension, we also provide an achievable rate region for the case where Bob applies independent decoding while Eve still employs joint decoding for detection.

$\bullet$ We extend the results of the DM MAC-WT channel to a MIMO Gaussian MAC-WT scenario, where we also consider two different decoding schemes for Bob.
Furthermore, we show that the analogous achievable rate region of a SISO Gaussian MAC-WT channel given in \cite[Theorem 1]{tekin2008general} does not hold in general. 
Though it gives a larger achievable rate region, we show in Appendix~\ref{Appendix_A} that there exist rate tuples in the region that are not achievable using the coding scheme given in \cite{tekin2008general}.
In this sense, our result provides a general achievable rate region for the MAC-WT scenario with confidential and open messages while \cite{tekin2008general} does not.

$\bullet$ Based on the information theoretic results, we then maximize the sum secrecy rate of the GV MAC-WT system by designing precoders for all users.
In particular, we first consider the problem for the case where Bob and Eve both jointly decode their interested messages.
As mentioned above, a similar optimization problem but with a special constraint has been studied in \cite{lee2017precoder}.
Note that the sum secrecy rate in \cite{lee2017precoder} is obtained based on \cite[Theorem 1]{tekin2008general}.
Though we show that the result in \cite[Theorem 1]{tekin2008general} unfortunately is not correct, the bound to the sum secrecy rate in \cite[Theorem 1]{tekin2008general} is the same as ours.
Then, we consider the sum secrecy rate maximization problem for the worse case where Bob independently decodes messages.
Since the problem is non-convex, we provide an iterative algorithm to obtain a suboptimal solution.

$\bullet$ The performance of the proposed algorithms is evaluated by simulation and compared with the exhaustive searching (ES) method as well as the scheme proposed in \cite{lee2017precoder}.
Simulation results show that the proposed algorithms perform very close to ES but involve much less computational complexity.
In contrast to the scheme given in \cite{lee2017precoder}, secure communication can be greatly enhanced since \cite{lee2017precoder} considered a special power constraint.
In addition, it is also shown that though we focus on the secrecy performance of the system, open messages can be simultaneously transmitted at a relatively high rate, which helps improve the system spectrum efficiency.

The rest of this paper is organized as follows. In Section~II, the general DM MAC-WT channel model is first given and achievable rate regions for the channel are then provided. In Section~III, the results are extended to a GV MAC-WT channel. In Section~VI, sum secrecy rate maximization problems are studied for the GV MAC-WT channel. Simulation results are presented in Section~V before conclusions in Section~IV.

Notations: we use calligraphic capital letters to denote sets, $|\cdot|$ to denote the cardinality of a set, `` $\setminus$ " to represent the set subtraction operation, and ${\cal X}_1 \times {\cal X}_2$ for the Cartesian product of the sets ${\cal X}_1$ and ${\cal X}_2$.
We use calligraphic subscript to denote the set of elements whose indexes take values from the subscript set, e.g., ${\cal X}_{\cal K} = \{{\cal X}_k, \forall k \in {\cal K}\}$, $X_{\cal K} = \{X_k, \forall k \in {\cal K}\}$, $\bm X_{\cal K} = \{\bm X_k, \forall k \in {\cal K}\}$, etc.
$\mathbb R$ and $\mathbb C$ are the real and complex spaces, respectively. 
Boldface upper (lower) case letters are used to denote matrices (vectors). 
A similar convention but with boldface upper-case letters is used for random vectors.
${\bm I}_B$ stands for the $B \times B$ dimensional identity matrix and $\bm 0$ denotes the all-zero vector or matrix.
Superscript $(\cdot)^H$ denotes the conjugated-transpose operation, ${\mathbb E}\left[\cdot\right]$ denotes the expectation operation, and $[\cdot]^+ \triangleq \max (\cdot,0)$.
The logarithm function $\log$ is base $2$.


\section{DM MAC-WT Channel}
\label{section3}

In this section, we first give the general DM MAC-WT channel model and define two metrics based on which coding schemes can be designed to guarantee perfect secrecy.
Then, we give two achievable rate regions for different decoding schemes at Bob.

\subsection{DM MAC-WT Channel Model}
\label{DM_MAC_channel_model}

\begin{figure}
	\centering
	\includegraphics[scale=0.40]{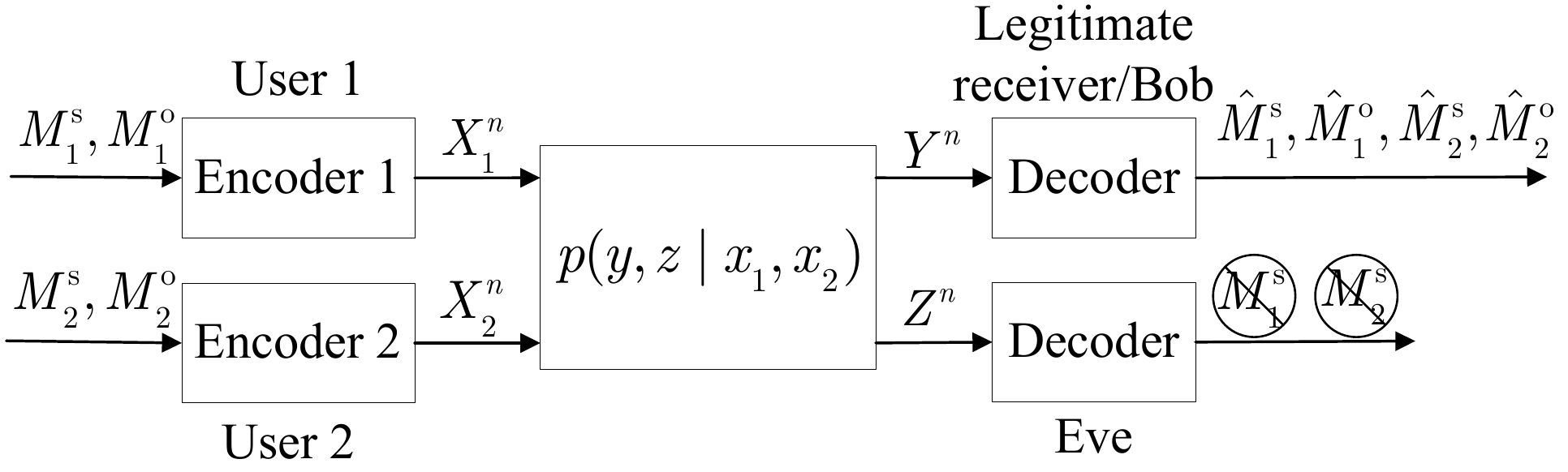}
	\caption{Block diagram of a two-user MAC-WT channel.}
	\label{Fig1}
\end{figure}

Consider a DM MAC-WT channel with $K$ users, a legitimate receiver, and an eavesdropper.
The block diagram of a simple two-user case is depicted in Fig.~\ref{Fig1}.
Let ${\cal K} = \{1, \cdots, K\}$ denote the set of all users.
The DM MAC-WT system can then be denoted by $\left({\cal X}_{\cal K}, p(y,z|x_{\cal K}), {\cal Y}, {\cal Z}\right)$ (in short $p(y,z|x_{\cal K})$), where ${\cal X}_k$, $\cal Y$, and $\cal Z$ are finite alphabets, $x_k \in {\cal X}_k$ is the channel input from user $k$, and $y \in {\cal Y}$ and $z \in {\cal Z}$ are respectively channel outputs at Bob and Eve. 
For brevity, we use the short-hand notation $p(x_k)$ to indicate $P_{X_k}(x_k)$ with $x_k \in {\cal X}_k$. 
Analogous short-hand notations are clear from the context. 

Each user $k \in {\cal K}$ wishes to communicate a secret message $M_k^{\text s}$ and an open message $M_k^{\text o}$ to Bob.
Hence, user $k$ encodes its messages into a codeword $X_k^n$ and transmits $X_k^n$ over channel $p(y,z|x_{\cal K})$.
Upon receiving the noisy sequence $Y^n$, Bob will decode the messages of all users.
To avoid leakage of confidential information to Eve, the secret messages of all users should be protected.
Let $R_k^{\text s}$ and $R_k^{\text o}$ denote the rate of user $k$'s secret and open messages, respectively. Then, a $\left( 2^{n R_1^{\text s}}, 2^{n R_1^{\text o}},  \cdots, 2^{n R_K^{\text s}}, 2^{n R_K^{\text o}}, n \right)$ secrecy code for the considered DM MAC-WT channel consists of
\begin{itemize}
	\item Secret and open message sets: ${\cal M}_k^{\text s} = \left[1:2^{n R_k^{\text s}}\right]$ and ${\cal M}_k^{\text o} = \left[1:2^{n R_k^{\text o}}\right], \forall k \in {\cal K}$. 
	Messages $M_k^{\text s}$ and $M_k^{\text o}$ are uniformly distributed over ${\cal M}_k^{\text s}$ and ${\cal M}_k^{\text o}$, respectively. 
	\item $K$ randomized encoders: the encoder of user $k$ maps message pair $(m_k^{\text s}, m_k^{\text o}) \in {\cal M}_k^{\text s} \times {\cal M}_k^{\text o}$ to a codeword $x_k^n$.
	\item A decoder at Bob which maps the received noisy sequence $y^n$ to message pairs $\left( {\hat m}_k^{\text s}, {\hat m}_k^{\text o} \right) \in  {\cal M}_k^{\text s} \times {\cal M}_k^{\text o}, \forall k \in {\cal K}$.
\end{itemize}

The average probability of error for a $\left( 2^{n R_1^{\text s}}, 2^{n R_1^{\text o}}, \cdots, \right.$ $\left. 2^{n R_K^{\text s}},2^{n R_K^{\text o}}, n \right)$ code is defined as
\begin{equation}\label{Pe}
P_{\text e} = P \left\{ \left( {\hat M}_{\cal K}^{\text s}, {\hat M}_{\cal K}^{\text o} \right) \neq \left( M_{\cal K}^{\text s}, M_{\cal K}^{\text o} \right) \right\}.
\end{equation}
The secrecy level of the MAC-WT system is evaluated by the information leakage rate at Eve, which is defined as
\begin{equation}\label{leakage_rate}
R_{{\text E}, {\cal S}} = \frac{1}{n} I (M_{\cal S}^{\text s}; Z^n), ~\forall~ {\cal S} \subseteq {\cal K}.
\end{equation}
For perfect secrecy of all transmitted secret messages, we would like $R_{{\text E}, {\cal S}} \rightarrow 0, \forall {\cal S} \subseteq {\cal K}$.
Since messages $M_k^{\text s}, \forall k \in {\cal K}$ are independent of each other, we have
\begin{align}\label{M1M2}
I (M_{\cal K}^{\text s}; Z^n) & = I (M_{\cal S}^{\text s}; Z^n) + I (M_{\bar {\cal S}}^{\text s}; Z^n | M_{\cal S}^{\text s})\nonumber\\
& \geq I (M_{\cal S}^{\text s}; Z^n) + I (M_{\bar {\cal S}}^{\text s}; Z^n), ~\forall~ {\cal S} \subseteq {\cal K},
\end{align}
where $\bar {\cal S}$ is the complementary set of $\cal S$, i.e., $\bar {\cal S} = \cal K \setminus \cal S$.
(\ref{M1M2}) indicates that if the leakage rate for all confidential messages vanishes, then the system is secure also for all possible message subsets.
Then, a rate tuple $(R_1^{\text s}, R_1^{\text o}, \cdots, R_K^{\text s}, R_K^{\text o})$ is said to be achievable if for any $\delta > 0$ there exists a sequence of 
$\left( 2^{n R_1^{\text s}}, 2^{n R_1^{\text o}}, \cdots, 2^{n R_K^{\text s}}, 2^{n R_K^{\text o}}, n \right)$ codes such that
\begin{align}
& \lim_{n \rightarrow \infty} P_{\text e} \leq \delta,\\
& \lim_{n \rightarrow \infty} R_{{\text E}, {\cal K}} \leq \delta.\label{RE}
\end{align}

\subsection{Main Results}
\label{main_results}

In the following we give an achievable rate region for the considered DM MAC-WT channel.
For the sake of convenience, we first give the results for the two-user case and then extend to the more general $K$-user case.

\begin{theorem}\label{theorem1}
	When $K=2$ and $(X_1, X_2, Y, Z) \sim $ $p(x_1) p(x_2) p(y,z| x_1, x_2)$, any rate tuple $\left(R_1^{\text s}, R_1^{\text o}, R_2^{\text s}, R_2^{\text o}\right)$ satisfying
	\begin{equation}\label{rate_region}
	\!\left\{\!\!\!
	\begin{array}{ll}
	\sum\limits_{k \in {\cal S}} (R_k^{\text s} + R_k^{\text o}) \leq I(X_{\cal S}; Y| X_{\bar {\cal S}}), \forall \cal S \subseteq \cal K, \\
	\sum\limits_{k \in {\cal S}} R_k^{\text s} \leq \left[I(X_{\cal S}; Y| X_{\bar {\cal S}}) - I(X_{\cal S}; Z) \right]^+\!\!, \forall \cal S \subseteq \cal K,\\
	R_1^{\text s} + R_2^{\text s} + R_k^{\text o} \leq \left[I(X_{\cal K}; Y) - I(X_{\bar k}; Z) \right]^+\!\!, \forall k \in \cal K,
	\end{array} \right.\!\!
	\end{equation}
	is achievable, where ${\bar k} = 1$ if $k=2$, and ${\bar k} = 2$ if $k=1$.
	Let ${\mathscr R} (X_1, X_2)$ denote the set of rate tuples satisfying (\ref{rate_region}).
	Then, the convex hull of the union of ${\mathscr R} (X_1, X_2)$ over all $p(x_1)  p(x_2)$ is an achievable rate region of the considered DM MAC-WT channel.
\end{theorem}

\itshape \textbf{Proof:}  \upshape
	See Appendix \ref{Prove_theorem1}.
\hfill $\Box$

The result in Theorem \ref{theorem1} can be directly extended to the more general case with $K \geq 1$ users.
\begin{lemma}\label{lemma1}
	For a general $K \geq 1$, let $(X_{\cal K}, Y, Z) $ $\sim \prod_{k=1}^K p(x_k) p(y,z| x_{\cal K})$. 
	Then, any rate tuple $\left(R_1^{\text s}, R_1^{\text o}, \cdots,\right.$ $\left. R_K^{\text s}, R_K^{\text o}\right)$ satisfying
	\begin{align}\label{rate_region0}
	\sum_{k \in \cal S} R_k^{\text s} + \sum_{k \in {\cal S}\setminus {\cal S}_1} R_k^{\text o} \leq & \left[I(X_{\cal S}; Y| X_{\bar {\cal S}}) - I(X_{{\cal S}_1}; Z) \right]^+, \nonumber\\
	& \forall {\cal S} \subseteq {\cal K} ~{\text {and}}~ {{\cal S}_1} \subseteq \cal S,
	\end{align}
	is achievable.
	Let ${\mathscr R} (X_{\cal K})$ denote the set of rate tuples satisfying (\ref{rate_region0}).
	Then, the convex hull of the union of ${\mathscr R} (X_{\cal K})$ over all $\prod_{k=1}^K p(x_k)$ is an achievable rate region of the DM MAC-WT channel with $K$ users.
\end{lemma}
\itshape \textbf{Proof:}  \upshape
This lemma can be proven by a simple extension of the proof of Theorem \ref{theorem1}.
\hfill $\Box$

To achieve the rate tuples in Lemma~\ref{lemma1}, Bob has to apply the joint typicality decoding scheme.
Since the complexity of joint decoding grows exponentially with the signal dimension, we also provide an achievable rate region in the following lemma for the case where Bob independently decodes the messages.
As for Eve, the most powerful joint decoding is still considered since the system secrecy may break down if we design the system by assuming a weaker Eve while it actually makes use of an enhanced receiver.
In contrast to the case where Bob uses joint decoding, this obviously causes a worse case from a system secrecy performance perspective.

\begin{lemma}\label{lemma01}
	When $K \geq 1$ and $(X_{\cal K}, Y, Z) $ $\sim \prod_{k=1}^K p(x_k) p(y,z| x_{\cal K})$, if Bob independently decodes the information of all users, while there is no restriction on the decoding scheme at Eve, then, any rate tuple $\left(R_1^{\text s}, R_1^{\text o}, R_2^{\text s}, R_2^{\text o}\right)$ satisfying
	\begin{equation}\label{rate_region01}
	\left\{\!\!\!
	\begin{array}{ll}
	R_k^{\text s} + R_k^{\text o} \leq I(X_k; Y), \forall k \in \cal K, \\
	\sum\limits_{k \in {\cal S}} R_k^{\text s} \leq \left[\sum\limits_{k \in {\cal S}} I(X_k; Y) - I(X_{\cal S}; Z) \right]^+\!\!, \forall \cal S \subseteq \cal K,
	\end{array} \right.\!\!
	\end{equation}
	is achievable.
	Let ${\hat {\mathscr R}} (X_{\cal K})$ denote the set of rate tuples satisfying (\ref{rate_region01}).
	Then, the convex hull of the union of ${\hat {\mathscr R}} (X_{\cal K})$ over all $\prod_{k=1}^K p(x_k)$ is an achievable rate region of the DM MAC-WT channel with $K$ users when Bob applies independent decoding.
\end{lemma}	

\itshape \textbf{Proof:}  \upshape
See Appendix \ref{Prove_lemma01}.
\hfill $\Box$

When Bob jointly decodes messages, let ${\cal S} = {\cal K}$ and ${\cal S}_1 = \phi$ in (\ref{rate_region0}), where $\phi$ is the empty set.
The maximum achievable sum rate $\sum_{k=1}^K \left( R_k^{\text s} + R_k^{\text o} \right)$ can then be obtained and denoted by
\begin{equation}\label{R_joint_DM}
R_{\text {joint}} (X_{\cal K}) = I(X_{\cal K}; Y).
\end{equation}
In addition, by setting ${\cal S} = {\cal S}_1 = {\cal K}$ in (\ref{rate_region0}), the maximum achievable sum secrecy rate $\sum_{k=1}^K R_k^{\text s}$ is given by
\begin{equation}\label{R_s_joint_DM}
R_{\text {joint}}^{\text s} (X_{\cal K}) = \left[I(X_{\cal K}; Y) - I(X_{\cal K}; Z)\right]^+.
\end{equation}
Similarly, when Bob independently decodes messages, using Lemma~\ref{lemma01}, we can get the maximum achievable sum rate $\sum_{k=1}^K \left( R_k^{\text s} + R_k^{\text o} \right)$ and sum secrecy rate $\sum_{k=1}^K R_k^{\text s}$
\begin{align}
R_{\text {inde}} (X_{\cal K}) & = \sum\limits_{k \in {\cal K}} I(X_k; Y), \label{R_inde_DM}\\
R_{\text {inde}}^{\text s} (X_{\cal K}) & = \left[\sum\limits_{k \in {\cal K}} I(X_k; Y) - I(X_{\cal K}; Z)\right]^+.\label{R_s_inde_DM}
\end{align}
Since a wiretap channel is considered in this paper, we are especially concerned about the secrecy performance of the system.
Then, an interesting question is if the users encode their confidential messages at the maximum sum secrecy rate (\ref{R_s_joint_DM}) or (\ref{R_s_inde_DM}) (depending on the decoding scheme at Bob), what is the maximum sum rate at which the users could encode their open messages.
We give the answer in the following lemma.
For convenience, we assume that $I(X_{\cal K}; Y) - I(X_{\cal K}; Z) > 0$ and $\sum_{k \in {\cal K}} I(X_k; Y) - I(X_{\cal K}; Z) > 0$, since otherwise we have $R_k^{\text s} = 0, \forall k \in {\cal K}$, i.e., the system reduces to a normal MAC channel with only open messages.
\begin{lemma}\label{max_R_o}
	If the users transmit their confidential messages at the maximum sum secrecy rate (\ref{R_s_joint_DM}) or (\ref{R_s_inde_DM}) (depending on the decoding scheme at Bob), the maximum achievable sum rate at which the users could send their open messages is\
	\begin{equation}\label{R_o_DM}
	R_{\text {joint}}^{\text o} (X_{\cal K}) = R_{\text {inde}}^{\text o} (X_{\cal K}) = I(X_{\cal K}; Z),
	\end{equation}
	which is the difference of (\ref{R_joint_DM}) and (\ref{R_s_joint_DM}), or the difference of (\ref{R_inde_DM}) and (\ref{R_s_inde_DM}).
\end{lemma}
\itshape \textbf{Proof:}  \upshape
See Appendix \ref{Prove_max_R_o}.
\hfill $\Box$

Lemma~\ref{max_R_o} shows that in contrast to the works which only considered secret messages, the channel can be made fully use of since open messages can be sent at rate $I(X_{\cal K}; Z)$ even when we focus on the secrecy performance of the system.


\section{GV MAC-WT Channel}
\label{GV_MAC-WT}

\newcounter{TempEqCnt}
\setcounter{TempEqCnt}{\value{equation}}
\setcounter{equation}{19}
\begin{figure*}[hb]
	\hrulefill
	\begin{equation}\label{rate_region4}
	\left\{\!\!\!
	\begin{array}{ll}
	R_k^{\text s} + R_k^{\text o} \leq \log \det \left( \bm H_k \bm F_k \bm H_k^H \bm D_k^{-1} + \bm I_B \right), ~\forall~ k \in \cal K, \\
	\sum\limits_{k \in {\cal S}} R_k^{\text s} \leq \left[\! \sum\limits_{k \in {\cal S}} \log\! \det\! \left(\! \bm H_k \bm F_k \bm H_k^H \bm D_k^{-1} \!+\! \bm I_B \!\right) \!-\! \log\! \det\! \left(\! \sum\limits_{k \in {\cal S}} \!\bm G_k \bm F_k \bm G_k^H \!\!\left(\! \sum\limits_{j \in {\cal K} \setminus {\cal S}} \!\bm G_j \bm F_j \bm G_j^H \!+\! \sigma_E^2 \bm I_E \!\right)^{\!-1} \!\!\!+\! \bm I_E\! \right)\!\right]^+\!\!, \forall \cal S \subseteq \cal K,
	\end{array} \right.\!\!\!\!
	\end{equation}
\end{figure*}

In this section, we consider a GV MAC-WT channel and extend the results obtained in the previous section to the Gaussian MIMO case.

Assume that each user $k$, Bob, and Eve are respectively equipped with $T_k$, $B$, and $E$ antennas.
The output of the channel corresponding to the input ${\bm X}_k \in {\mathbb C}^{T_k \times 1}, \forall k \in \cal K$ is
\setcounter{equation}{13}
\begin{align}\label{GV_YZ}
& \bm Y = \sum_{k=1}^K \bm H_k \bm X_k + \bm N_{\text B},\nonumber\\
& \bm Z = \sum_{k=1}^K \bm G_k \bm X_k + \bm N_{\text E},
\end{align}
where $\bm H_k \in {\mathbb C}^{B \times T_k}$ and $\bm G_k \in {\mathbb C}^{E \times T_k}$ are constant channel gain matrices from user $k$ to Bob and Eve, and $\bm Y \in {\mathbb C}^{B \times 1}$, $\bm Z \in {\mathbb C}^{E \times 1}$, $\bm N_{\text B} \in {\mathbb C}^{B \times 1}$, and $\bm N_{\text E} \in {\mathbb C}^{B \times 1}$ are the received vectors as well as additive Gaussian noise vectors at Bob and Eve.
We consider discrete-time channel and assume that for every transmission $i \in [1,n]$, the noise vector processes $\left\{ \bm N_{{\text B}i} \right\}$ and  $\left\{ \bm N_{{\text E}i} \right\}$ are independent and identically distributed (i.i.d.) with $\bm N_{{\text B}i} \sim {\cal CN}(0, \sigma_B^2 \bm I_B)$ and $\bm N_{{\text E}i} \sim {\cal CN}(0, \sigma_E^2 \bm I_E)$.
Besides, for every codeword $\bm x_k^n = (\bm x_{k1}, \cdots, \bm x_{kn})$, we assume the following average transmission power constraint
\begin{equation}\label{power_constraint}
\sum_{i=1}^n \bm x_{ki}^H \bm x_{ki} \leq n P_k, ~\forall~ k \in \cal K.
\end{equation}

Define positive semi-definite matrices $\bm F_k \in {\mathbb C}^{T_k \times T_k}, \forall k \in \cal K$, and denote $\bm F_{\cal K} = \left\{ \bm F_k, \cdots, \bm F_K \right\}$ and ${\cal F} = \left\{ \bm F_{\cal K}: \bm F_k \succeq \bm 0,\right.$ $\left. {\text {tr}}(\bm F_k) \leq P_k, \forall k \in \cal K \right\}$.
In the following theorem, we give an achievable rate region for the GV MAC-WT channel when input $\bm X_k, ~\forall~ k \in \cal K$ are Gaussian vectors.
\begin{theorem}\label{theorem_GV_joint}
	If $\bm X_k \sim {\cal CN}(\bm 0, \bm F_k), ~\forall~ k \in \cal K$, $\bm F_{\cal K} \in {\cal F}$, any rate tuple $\left(R_1^{\text s}, R_1^{\text o}, \cdots, R_K^{\text s}, R_K^{\text o}\right)$ satisfying
	\begin{align}\label{rate_region3}
	& \sum_{k \in \cal S} R_k^{\text s} + \sum_{k \in {\cal S}\setminus {\cal S}_1} R_k^{\text o} \leq \left[I(\bm X_{\cal S}; \bm Y| \bm X_{\bar {\cal S}}) - I(\bm X_{{\cal S}_1}; \bm Z) \right]^+ \nonumber\\
	& = \left[ \log \det \left( \sum_{k \in {\cal S}} \frac{1}{\sigma_B^2} \bm H_k \bm F_k \bm H_k^H + \bm I_B \right) \right. \nonumber\\
	& \left.- \log\! \det\!\! \left(\! \sum_{k \in {\cal S}_1} \!\!\bm G_k \bm F_k \bm G_k^H \!\!\left(\! \sum_{j \in {\cal K} \setminus {\cal S}_1} \!\!\!\bm G_j \bm F_j \bm G_j^H \!+\! \sigma_E^2 \bm I_E \!\right)^{-1} \!\!\!+\! \bm I_E\! \right)\!\right]^+, \nonumber\\
	&~\forall~ {{\cal S} \subseteq {\cal K}}, ~{\text {and}}~ {{\cal S}_1 \subseteq \cal S},
	\end{align}
	is achievable.
	Let ${\mathscr R}_{\text G} (\bm F_{\cal K})$ denote the set of rate tuples satisfying (\ref{rate_region3}) with subscript `G' to indicate the Gaussian channel.
	Then, the convex hull of the union of ${\mathscr R}_{\text G} (\bm F_{\cal K})$ over all $\bm F_{\cal K} \in {\cal F}$ is an achievable rate region of the considered GV MAC-WT channel with Gaussian input.
\end{theorem}
\itshape \textbf{Proof:}  \upshape
See Appendix \ref{prove_lemma3}.
\hfill $\Box$

\begin{remark}\label{remark1}
	In reference \cite{tekin2008general}, a SISO Gaussian MAC-WT channel is considered.
	A superposition encoding rate region, in which the rate tuples $\left(R_1^{\text s}, R_1^{\text o}, \cdots, R_K^{\text s}, R_K^{\text o}\right)$ satisfy
	\begin{equation}\label{rate_region11}
	\left\{\!\!\!
	\begin{array}{ll}
	\sum\limits_{k \in {\cal S}} (R_k^{\text s} + R_k^{\text o}) \leq I(X_{\cal S}; Y| X_{\bar {\cal S}}), \forall \cal S \subseteq \cal K, \\
	\sum\limits_{k \in {\cal S}} R_k^{\text s} \leq \left[I(X_{\cal S}; Y| X_{\bar {\cal S}}) - I(X_{\cal S}; Z) \right]^+\!\!, \forall \cal S \subseteq \cal K,
	\end{array} \right.\!\!
	\end{equation}
	is given in \cite[eq.~(19)]{tekin2008general}. 
	Then, it is stated in \cite[Theorem 1]{tekin2008general} that the convex hull of the superposition encoding rate region union over all power constraint is achievable.
	In Appendix \ref{Appendix_A}, we show that the result in \cite[Theorem 1]{tekin2008general} unfortunately is not correct.
	In this sense, our result provides a general achievable rate region for the MAC-WT scenario with confidential and open messages while \cite[Theorem 1]{tekin2008general} does not.
\end{remark}

Similar to Lemma~\ref{lemma01}, we also give in the following lemma an achievable rate region for the GV MAC-WT channel with an independent-decoding Bob, which is sometimes more appealing in terms of practicality.

\begin{lemma}\label{lemma4}
	If $\bm X_k \sim {\cal CN}(\bm 0, \bm F_k), ~\forall~ k \in \cal K$, $\bm F_{\cal K} \in {\cal F}$, and Bob independently decodes users' messages, then, any rate tuple $\left(R_1^{\text s}, R_1^{\text o}, \cdots, R_K^{\text s}, R_K^{\text o}\right)$ satisfying 
	\begin{equation}\label{rate_region_GV_joint}
	\left\{\!\!\!
	\begin{array}{ll}
	R_k^{\text s} + R_k^{\text o} \leq I(\bm X_k; \bm Y), \forall k \in \cal K, \\
	\sum\limits_{k \in {\cal S}} R_k^{\text s} \leq \left[\sum\limits_{k \in {\cal S}} I(\bm X_k; \bm Y) - I(\bm X_{\cal S}; \bm Z) \right]^+\!\!, \forall \cal S \subseteq \cal K,
	\end{array} \right.\!\!
	\end{equation}
	is achievable.
	Denoting 
	\begin{equation}\label{D_k}
	\bm D_k = \sum_{j \in {\cal K} \setminus k} \bm H_j \bm F_j \bm H_j^H + \sigma_B^2 \bm I_B,
	\end{equation}
	(\ref{rate_region_GV_joint}) can be rewritten as (\ref{rate_region4}) at the bottom of this page.
	Let ${\hat {\mathscr R}}_{\text G} (\bm F_{\cal K})$ denote the set of rate tuples satisfying (\ref{rate_region_GV_joint}) or (\ref{rate_region4}).
	Then, the convex hull of the union of ${\hat {\mathscr R}}_{\text G} (\bm F_{\cal K})$ over all $\bm F_{\cal K} \in {\cal F}$ is an achievable rate region of the GV MAC-WT channel with Gaussian input and independent-decoding Bob.
\end{lemma}

Analogous to (\ref{R_joint_DM}) $\sim$ (\ref{R_s_inde_DM}), when Bob applies different decoding schemes, we can get the following maximum achievable sum rate $\sum_{k=1}^K \left( R_k^{\text s} + R_k^{\text o} \right)$ and sum secrecy rate$\sum_{k=1}^K R_k^{\text s}$ for the considered GV MAC-WT channel
\setcounter{equation}{20}
\begin{align}
R_{\text {joint}} (\bm F_{\cal K}) & = I(\bm X_{\cal K}; \bm Y) \nonumber\\
& = \log \det \left( \sum_{k=1}^K \frac{1}{\sigma_B^2} \bm H_k \bm F_k \bm H_k^H + \bm I_B \right), \label{R_joint_GV}\\
R_{\text {joint}}^{\text s} (\bm F_{\cal K}) & = \left[I(\bm X_{\cal K}; \bm Y) - I(\bm X_{\cal K}; \bm Z)\right]^+ \nonumber\\
& = \left[\log \det \left( \sum_{k=1}^K \frac{1}{\sigma_B^2} \bm H_k \bm F_k \bm H_k^H + \bm I_B \right)\right.\nonumber\\
& \left. - \log \det \left( \sum_{k=1}^K \frac{1}{\sigma_E^2} \bm G_k \bm F_k \bm G_k^H + \bm I_E \right)\right]^+,\label{R_s_joint_GV}
\end{align}
and
\begin{align}
R_{\text {inde}} (\bm F_{\cal K}) & = \sum\limits_{k \in {\cal K}} I(\bm X_k; \bm Y) \nonumber\\
& = \sum_{k=1}^K \log \det \left( \bm H_k \bm F_k \bm H_k^H \bm D_k^{-1} + \bm I_B \right), \label{R_inde_GV}\\
R_{\text {inde}}^{\text s} (\bm F_{\cal K}) & = \left[\sum\limits_{k \in {\cal K}} I(\bm X_k; \bm Y) - I(\bm X_{\cal K}; \bm Z)\right]^+ \nonumber\\
& = \left[\sum_{k=1}^K \log \det \left( \bm H_k \bm F_k \bm H_k^H \bm D_k^{-1} + \bm I_B \right) \right.\nonumber\\
& \left. - \log \det \left( \sum_{k=1}^K \frac{1}{\sigma_E^2} \bm G_k \bm F_k \bm G_k^H + \bm I_E \right)\right]^+. \label{R_s_inde_GV}
\end{align}
In addition, we give the following lemma which can be similarly proven as Lemma~\ref{max_R_o}
\begin{lemma}\label{max_R_o_inde}
	If the users transmit their confidential messages at the maximum sum secrecy rate (\ref{R_s_joint_GV}) or (\ref{R_s_inde_GV}) (depending on the decoding scheme at Bob), the maximum achievable sum rate at which they could send their open messages is
	\begin{align}\label{R_o_GV}
	R_{\text {joint}}^{\text o} (\bm F_{\cal K}) & = R_{\text {inde}}^{\text o} (\bm F_{\cal K})\nonumber\\
	& = I(\bm X_{\cal K}; \bm Z)\nonumber\\
	& = \log \det \left( \sum_{k=1}^K \frac{1}{\sigma_E^2} \bm G_k \bm F_k \bm G_k^H + \bm I_E \right).
	\end{align}
	which is the difference of (\ref{R_joint_GV}) and (\ref{R_s_joint_GV}), or the difference of (\ref{R_inde_GV}) and (\ref{R_s_inde_GV}).
\end{lemma}

Note that though $R_{\text {inde}}^{\text o} (\bm F_{\cal K})$ has the same formulation as $R_{\text {joint}}^{\text o} (\bm F_{\cal K})$ in (\ref{R_o_GV}), they may have different values since $\bm F_{\cal K}$ may be respectively obtained by solving different problems, e.g., (\ref{C1_max}) and (\ref{C2_max}) in the next section.
Since besides secret messages, the users may simultaneously transmit open messages, the spectrum efficiency of the GV MAC-WT system can be improved.

\section{Sum Secrecy Rate Maximization for the GV MAC-WT Channel}
\label{Sum_SR_max}

From Theorem~\ref{theorem_GV_joint}, Lemma~\ref{lemma4}, and (\ref{R_joint_GV}) $\sim$ (\ref{R_s_inde_GV}), it is known that for given channel gain matrices and power constraint, the bounds of regions ${\mathscr R}_{\text G} (\bm F_{\cal K})$ and ${\hat {\mathscr R}}_{\text G} (\bm F_{\cal K})$ depend on the covariance matrices of all users' Gaussian input vectors, i.e., $\bm F_k, \forall k \in {\cal K}$, or $\bm F_{\cal K}$ in short.
We are thus interested in maximizing the sum rate of the GV MAC-WT system by optimizing $\bm F_{\cal K}$.

The problem of maximizing $R_{\text {joint}} (\bm F_{\cal K})$ subject to the power constraint is the classical sum-capacity maximization problem for a GV MAC channel.
Since $R_{\text {joint}} (\bm F_{\cal K})$ is concave with respect to (w.r.t.) $\bm F_{\cal K}$, the problem is convex and the optimal solution can be efficiently obtained by using iterative water-filling method \cite[Remark 9.4]{el2011network}, \cite{1262622}.  
Moreover, the problem of maximizing the sum rate $R_{\text {inde}} (\bm F_{\cal K})$ in (\ref{R_inde_GV}) subject to the power constraint has been widely studied.
Due to the non-convexity, a sub-optimal solution of the problem can be obtained by using the weighted sum mean-square-error minimization (WMMSE) scheme proposed in \cite{shi2011iteratively}.  

As a result, we focus on the sum secrecy rate maximization problems in this section.
For convenience, we omit the $\left[ \cdot \right]^+$ operation in $R_{\text {joint}}^{\text s} (\bm F_{\cal K})$ and $R_{\text {inde}}^{\text s} (\bm F_{\cal K})$ when solving the corresponding problems and check whether they are positive or not once the problems are solved.

\subsection{Sum secrecy rate maximization when Bob jointly decodes messages}
\label{SSR_max_joint} 

When Bob jointly decodes messages, the sum secrecy rate problem can be formulated as
\begin{subequations}\label{C1_max}
	\begin{align}
	\mathop {\max }\limits_{\bm F_{\cal K}} \quad & R_{\text {joint}}^{\text s} (\bm F_{\cal K}) \label{C1_max_a}\\
	\text{s.t.} \quad\; &  {\text {tr}}(\bm F_k) \leq P_k, ~\forall~ k \in \cal K, \label{C1_max_b}\\
	& \bm F_k \succeq \bm 0, ~\forall~ k \in \cal K. \label{C1_max_c}
	\end{align}
\end{subequations}

Actually, a similar problem with the same objective function as (\ref{C1_max_a}) has been studied in \cite{lee2017precoder}.
Differently, instead of forcing a power constraint on the trace of $\bm F_k$ as in (\ref{C1_max_b}), \cite{lee2017precoder} considered constraint $\bm F_k \preceq \bm S_k$ for each user $k$, where $\bm S_k$ is a fixed positive semi-definite matrix and satisfies ${\text {tr}}(\bm S_k) \leq P_k$.
Obviously, the problem in \cite{lee2017precoder} is a special case of problem (\ref{C1_max}).
Note that the sum secrecy rate in \cite{lee2017precoder} is obtained based on \cite[Theorem 1]{tekin2008general}.
Though we have shown in Remark \ref{remark1} and Appendix \ref{Appendix_A} that the result in \cite[Theorem 1]{tekin2008general} unfortunately is not correct, the bound to the sum secrecy rate in \cite[Theorem 1]{tekin2008general} is the same as ours.

Before solving problem (\ref{C1_max}), we first consider an important special SIMO case with $T_k = 1, \forall k \in {\cal K}$.
In this case, (\ref{C1_max}) reduces to the following power control problem
\begin{subequations}\label{C1_max_SIMO}
	\begin{align}
	\mathop {\max }\limits_{\bm f} \quad & \log \det \left( \sum_{k=1}^K \frac{1}{\sigma_B^2} f_k \bm h_k \bm h_k^H + \bm I_B \right)\nonumber\\
	& - \log \det \left( \sum_{k=1}^K \frac{1}{\sigma_E^2} f_k \bm g_k \bm g_k^H + \bm I_E \right) \label{C1_max_SIMO_a}\\
	\text{s.t.} \quad\; &  0 \leq f_k \leq P_k, ~\forall~ k \in \cal K, \label{C1_max_SIMO_b}
	\end{align}
\end{subequations}
where $\bm h_k \in {\mathbb C}^{B \times 1}$ and $\bm g_k \in {\mathbb C}^{E \times 1}$ are the channel vectors, $f_k$ is the transmit power of user $k$, and $\bm f = (f_1, \cdots, f_K)^T$.
By checking the monotonicity of $R_{\text {joint}}^{\text s} (\bm f)$ w.r.t. $f_k$, it was proven in \cite[Theorem~2]{lee2017precoder} that the optimal solution of (\ref{C1_max_SIMO}) can be easily found.
For completeness, we restate \cite[Theorem~2]{lee2017precoder} in the following lemma.
\begin{lemma}\label{joing_SIMO}
	When $T_k = 1, \forall k \in {\cal K}$, the optimal solution of problem (\ref{C1_max_SIMO}) is obtained by either $f_k = P_k$ or $f_k = 0, \forall k \in {\cal K}$, i.e., each user $k$ either transmits in the maximum power or keeps inactive.
\end{lemma}
Based on Lemma~\ref{joing_SIMO}, the optimal solution of problem (\ref{C1_max_SIMO}) can be found by searching over $2^K$ possible solutions.

Since both $\log \det \left( \sum_{k=1}^K \frac{1}{\sigma_B^2} \bm H_k \bm F_k \bm H_k^H + \bm I_B \right)$ and $\log \det \left( \sum_{k=1}^K \frac{1}{\sigma_E^2} \bm G_k \bm F_k \bm G_k^H + \bm I_E \right)$ are concave functions, for the general MIMO case, problem (\ref{C1_max}) is a difference of convex (DC) programming.
A sub-optimal solution of problem (\ref{C1_max}) can thus be obtained by applying the iterative majorization minimization (MM) based algorithm, which solves a sequence of convex problems by linearizing the original non-convex functions in each iteration \cite{7547360, jeon2016joint}.
Let $\{\bm F_{k0}\}$ denote the solution obtained in the last iteration.
By utilizing the first-order Taylor series approximation \cite{deadman2016taylor}, an upper bound to $\log \det \left( \sum_{k=1}^K \frac{1}{\sigma_E^2} \bm G_k \bm F_k \bm G_k^H + \bm I_E \right)$ can be obtained as follows
\begin{align}\label{Taylor_joint}
& \log \det \left( \sum_{k=1}^K \frac{1}{\sigma_E^2} \bm G_k \bm F_{k0} \bm G_k^H + \bm I_E \right) \nonumber\\
& + \sum_{k=1}^K {\text {tr}} \left[ \bm G_k^H \bm A_0^{-1} \bm G_k \left( \bm F_k - \bm F_{k0} \right) \right],
\end{align}
where $\bm A_0 = \sum_{j=1}^K \bm G_j \bm F_{j0}^H \bm G_j^H + \sigma_E^2 \bm I_E$.
Ignoring the constant terms in (\ref{Taylor_joint}), a sub-optimal solution of problem (\ref{C1_max}) can then be obtained by iteratively solving
\begin{subequations}\label{C1_max2}
	\begin{align}
	\mathop {\max }\limits_{\bm F_{\cal K}} \quad & \log \det \left( \sum_{k=1}^K \frac{1}{\sigma_B^2} \bm H_k \bm F_k \bm H_k^H + \bm I_B \right)\nonumber\\
	& + \sum_{k=1}^K {\text {tr}} \left( \bm G_k^H \bm A_0^{-1} \bm G_k \bm F_k \right) \label{C1_max2_a}\\
	\text{s.t.} \quad\; &  {\text {tr}}(\bm F_k) \leq P_k, ~\forall~ k \in \cal K, \label{C1_max2_b}\\
	& \bm F_k \succeq \bm 0, ~\forall~ k \in \cal K. \label{C1_max2_c}
	\end{align}
\end{subequations}
Problem (\ref{C1_max2}) is convex and can thus be optimally solved by using general tools, e.g., CVX, etc.
The alternative algorithm for solving problem (\ref{C1_max}) is summarized in Algorithm~\ref{algorithm_joint}.
\begin{algorithm}[h]
	\begin{algorithmic}[1]
		\caption{Alternative algorithm for solving problem (\ref{C1_max})}
		\State Initialize $\bm F_k, ~\forall~ k \in \cal K$.
		\Repeat
		\State Let $\bm F_{k0} = \bm F_k, ~\forall~ k \in \cal K$.
		\State Update $\bm F_k, ~\forall~ k \in \cal K$ by solving (\ref{C1_max2}).
		\Until convergence
		\label{algorithm_joint}
	\end{algorithmic}
\end{algorithm}

\subsection{Sum secrecy rate maximization when Bob independently decodes messages}
\label{SSR_max_inde}

If Bob independently decodes the information of all users, while there is no restriction on the decoding scheme at Eve, the sum secrecy rate problem can be formulated as
\begin{subequations}\label{C2_max}
	\begin{align}
	\mathop {\max }\limits_{\bm F_{\cal K}} \quad & R_{\text {inde}}^{\text s} (\bm F_{\cal K}) \label{C2_max_a}\\
	\text{s.t.} \quad\; &  {\text {tr}}(\bm F_k) \leq P_k, ~\forall~ k \in \cal K, \label{C2_max_b}\\
	& \bm F_k \succeq \bm 0, ~\forall~ k \in \cal K. \label{C2_max_c}
	\end{align}
\end{subequations}
It can be readily verified that problem (\ref{C2_max}) is non-convex.
To solve this problem, we first consider the following GV MAC-WT channel model
\begin{align}\label{GV_YZ_hat}
& {\hat {\bm Y}} = \sum_{k=1}^K \bm H_k \bm V_k {\hat {\bm X}}_k + \bm N_{\text B},\nonumber\\
& {\hat {\bm Z}} = \sum_{k=1}^K \bm G_k \bm V_k {\hat {\bm X}}_k + \bm N_{\text E},
\end{align}
where ${\hat {\bm X}}_k \in {\mathbb C}^{T_k \times 1}$ is the transmit signal vector of user $k$, ${\hat {\bm X}}_k \sim {\cal CN}(0, \bm I_{T_k})$, and $\bm V_k \in {\mathbb C}^{T_k \times T_k}$ is the beamformer that user $k$ uses to transmit ${\hat {\bm X}}_k$.
The other parameters in (\ref{GV_YZ_hat}), e.g., $\bm H_k$, $\bm G_k$, $\bm N_{\text B}$, and $\bm N_{\text E}$, are the same as those in (\ref{GV_YZ}).
Then, by replacing $\bm F_k$ in (\ref{R_s_inde_GV}) with $\bm V_k \bm V_k^H$, the sum secrecy rate maximization problem for channel model (\ref{GV_YZ_hat}) is given by
\begin{subequations}\label{C3_max}
	\begin{align}
	\mathop {\max }\limits_{\bm V_{\cal K}} \quad & \sum_{k=1}^K \log \det \left( \bm H_k \bm V_k \bm V_k^H \bm H_k^H {\hat{\bm D}}_k^{-1} + \bm I_B \right)\nonumber\\
	& - \log \det \left( \sum_{k=1}^K \frac{1}{\sigma_E^2} \bm G_k \bm V_k \bm V_k^H \bm G_k^H + \bm I_E \right) \label{C3_max_a}\\
	\text{s.t.} \quad\; &  {\text {tr}}(\bm V_k \bm V_k^H) \leq P_k, ~\forall~ k \in \cal K, \label{C3_max_b}
	\end{align}
\end{subequations}
where ${\hat{\bm D}}_k = \sum_{j \in {\cal K} \setminus k} \bm H_j \bm V_j \bm V_j^H \bm H_j^H + \sigma_B^2 \bm I_B$.
Comparing (\ref{C2_max}) and (\ref{C3_max}), it can be easily found that any solution of problem (\ref{C3_max}) is also the solution of problem (\ref{C2_max}) by setting $\bm F_k = \bm V_k \bm V_k^H$, and any solution of problem (\ref{C2_max}) is also the solution of problem (\ref{C3_max}) by decomposing $\bm F_k$.
Channel model (\ref{GV_YZ_hat}) is thus equivalent to channel model (\ref{GV_YZ}) in terms of sum secrecy rate.
Hence, we can solve problem (\ref{C2_max}) by dealing with (\ref{C3_max}).
Though problem (\ref{C3_max}) is still non-convex, motivated by the WMMSE scheme proposed in \cite{shi2011iteratively}, we simplify the first term of objective function (\ref{C3_max_a}) as follows by introducing auxiliary variables
\begin{subequations}\label{problem1}
	\begin{align}
	\mathop {\min }\limits_{\bm V_{\cal K}, \bm W_{\cal K}, \bm U_{\cal K}} \quad & \sum_{k=1}^K \left[ {\text {tr}}(\bm U_k \bm J_k) - \log \det (\bm U_k) \right]\nonumber\\
	& + \log \det \left( \sum_{k=1}^K \frac{1}{\sigma_E^2} \bm G_k \bm V_k \bm V_k^H \bm G_k^H + \bm I_E \right) \label{problem1_a}\\
	\text{s.t.} \quad\quad\;\; &  {\text {tr}}(\bm V_k \bm V_k^H) \leq P_k, ~\forall~ k \in \cal K, \label{problem1_b}\\
	& \bm U_k \succ \bm 0, ~\forall~ k \in \cal K, \label{problem1_c}
	\end{align}
\end{subequations}
where $\bm W_k \in {\mathbb C}^{B \times T_k}$ can be seen as the MMSE receiver adopted by Bob for detecting ${\hat {\bm X}}_k$, $\bm U_k \in {\mathbb C}^{T_k \times T_k}$ is an auxiliary weight matrix variable, and $\bm J_k$ is the MSE matrix, i.e.,
\begin{align}\label{J_k}
\bm J_k & = {\mathbb E} \left[ \left( \bm W_k^H {\hat {\bm Y}} - {\hat {\bm X}}_k \right) \left( \bm W_k^H {\hat {\bm Y}} - {\hat {\bm X}}_k \right)^H  \right]\nonumber\\
& = \bm W_k^H \bm A_1 \bm W_k - \bm W_k^H \bm H_k \bm V_k - \bm V_k^H \bm H_k^H \bm W_k + \bm I_{T_k},
\end{align}
with $\bm A_1 = \sum_{j=1}^K \bm H_j \bm V_j \bm V_j^H \bm H_j^H + \sigma_B^2 \bm I_B$.
It can be proven as \cite[Theorem 1]{shi2011iteratively} that problem (\ref{problem1}) is equivalent to (\ref{C3_max}) in terms of sum secrecy rate. 
The objective function (\ref{problem1_a}) is convex with respect to $\bm W_{\cal K}$ and $\bm U_{\cal K}$.
Hence, we solve problem (\ref{problem1}) by optimizing variables $\bm W_{\cal K}$, $\bm U_{\cal K}$, and $\bm V_{\cal K}$ in an alternative manner.
First, for given $\bm U_{\cal K}$ and $\bm V_{\cal K}$, the optimal MMSE receivers can be obtained as follows 
\begin{align}\label{W_k}
\bm W_k^* & = \mathop {\arg \min}\limits_{\bm W_k} {\text {tr}}(\bm J_k) \nonumber\\
& = \bm A_1^{-1} \bm H_k \bm V_k, ~\forall~ k \in \cal K.
\end{align}
Second, for given $\bm W_{\cal K}$ and $\bm V_{\cal K}$, the optimal weight matrix variables are give by
\begin{align}\label{U_k}
\bm U_k^* & = \bm J_k^{-1} \nonumber\\
& \overset{(a)}{=} \left( \bm I_{T_k} - \bm V_k^H \bm H_k^H \bm A_1^{-1} \bm H_k \bm V_k \right)^{-1} \nonumber\\
& \overset{(b)}{=} \bm I_{T_k} + \bm V_k^H \bm H_k^H {\hat{\bm D}}_k^{-1} \bm H_k \bm V_k, ~\forall~ k \in \cal K,
\end{align}
where $(a)$ is obtained by substituting (\ref{W_k}) into (\ref{J_k}) and $(b)$ follows from applying the Woodbury matrix identity.
Then, we need to optimize $\bm V_{\cal K}$ for given $\bm W_{\cal K}$ and $\bm U_{\cal K}$, i.e., solving problem 
\begin{subequations}\label{problem2}
	\begin{align}
	\mathop {\min }\limits_{\bm V_{\cal K}} \quad & \sum_{k=1}^K {\text {tr}}(\bm U_k \bm J_k) + \log \det \left( \sum_{k=1}^K \frac{1}{\sigma_E^2} \bm G_k \bm V_k \bm V_k^H \bm G_k^H + \bm I_E \right) \label{problem2_a}\\
	\text{s.t.} \quad\; &  {\text {tr}}(\bm V_k \bm V_k^H) \leq P_k, ~\forall~ k \in \cal K. \label{problem2_b}
	\end{align}
\end{subequations}
Though $\sum_{k=1}^K {\text {tr}}(\bm U_k \bm J_k)$ is convex w.r.t. $\bm V_{\cal K}$, problem (\ref{problem2}) is non-convex since $\log \det \left( \sum_{k=1}^K \frac{1}{\sigma_E^2} \bm G_k \bm V_k \bm V_k^H \bm G_k^H + \bm I_E \right)$ is non-convex.
To solve this problem, we replace $\bm V_k \bm V_k^H$ in $\log \det \left( \sum_{k=1}^K \frac{1}{\sigma_E^2} \bm G_k \bm V_k \bm V_k^H \bm G_k^H + \bm I_E \right)$ with $\bm F_k$ and get the following equivalent problem
\begin{subequations}\label{problem3}
	\begin{align}
	\mathop {\min }\limits_{\bm V_{\cal K}, \bm F_{\cal K}} \quad & \sum_{k=1}^K {\text {tr}}(\bm U_k \bm J_k) + \log \det \left( \sum_{k=1}^K \frac{1}{\sigma_E^2} \bm G_k \bm F_k \bm G_k^H + \bm I_E \right) \label{problem3_a}\\
	\text{s.t.} \quad\;\; &  {\text {tr}}(\bm V_k \bm V_k^H) \leq P_k, ~\forall~ k \in \cal K, \label{problem3_b}\\
	&  \bm F_k = \bm V_k \bm V_k^H, ~\forall~ k \in \cal K. \label{problem3_c}
	\end{align}
\end{subequations}
Since $\log \det \left( \sum_{k=1}^K \frac{1}{\sigma_E^2} \bm G_k \bm F_k \bm G_k^H + \bm I_E \right)$ is concave in $\bm F_{\cal K}$, (\ref{problem3}) is a DC programming.
We can thus handle problem (\ref{problem3}) by applying the iterative MM algorithm.
In particular, as in the previous subsection, we first introduce $\{\bm F_{k0}\}$, which is the solution obtained in the last iteration, and get (\ref{Taylor_joint}), which is an upper bound to $\log \det \left( \sum_{k=1}^K \frac{1}{\sigma_E^2} \bm G_k \bm F_k \bm G_k^H + \bm I_E \right)$.
Then, ignoring the constant terms in (\ref{Taylor_joint}), a sub-optimal solution of problem (\ref{problem3}) can be obtained by iteratively solving
\begin{subequations}\label{problem4}
	\begin{align}
	\mathop {\min }\limits_{\bm V_{\cal K}, \bm F_{\cal K}} \quad & \sum_{k=1}^K {\text {tr}}(\bm U_k \bm J_k) + \sum_{k=1}^K {\text {tr}} \left( \bm G_k^H \bm A_0^{-1} \bm G_k \bm F_k \right) \label{problem4_a}\\
	\text{s.t.} \quad\;\; &  {\text {tr}}(\bm V_k \bm V_k^H) \leq P_k, ~\forall~ k \in \cal K, \label{problem4_b}\\
	&  \bm F_k = \bm V_k \bm V_k^H, ~\forall~ k \in \cal K, \label{problem4_c}
	\end{align}
\end{subequations}
and the convergence of this iterative process is guaranteed.
Obviously, problem (\ref{problem4}) is equivalent to
\begin{subequations}\label{problem5}
	\begin{align}
	\mathop {\min }\limits_{\bm V_{\cal K}} \quad & \sum_{k=1}^K {\text {tr}}(\bm U_k \bm J_k) + \sum_{k=1}^K {\text {tr}} \left( \bm G_k^H \bm A_0^{-1} \bm G_k \bm V_k \bm V_k^H \right) \label{problem5_a}\\
	\text{s.t.} \quad\; &  {\text {tr}}(\bm V_k \bm V_k^H) \leq P_k, ~\forall~ k \in \cal K, \label{problem5_b}
	\end{align}
\end{subequations}
which is a convex quadratic optimization problem and can be optimally solved.
Attaching a Lagrange multiplier $\lambda_k$ to the power constraint of each user $k$, we get the following Lagrange function
\begin{align}\label{Lagrange}
& {\cal L} \left( \bm V_{\cal K}, \lambda_{\cal K} \right) \nonumber\\
= & \sum_{k=1}^K {\text {tr}} \left\{ \sum_{j=1}^K \bm U_j \bm W_j^H \bm H_k \bm V_k \bm V_k^H \bm H_k^H \bm W_j - \bm U_k \bm W_k^H \bm H_k \bm V_k \right. \nonumber\\
- & \bm U_k \bm V_k^H \bm H_k^H \bm W_k + \bm U_k + \bm G_k^H \bm A_0^{-1} \bm G_k \bm V_k \bm V_k^H + \lambda_k \bm V_k \bm V_k^H  \Bigg\}\nonumber\\
- & \sum_{k=1}^K \lambda_k P_k.
\end{align}
The first-order optimality condition of ${\cal L} \left( \bm V_{\cal K}, \lambda_{\cal K} \right)$ with respect to $\bm V_k$ yields
\begin{align}\label{V_k}
& \bm V_k (\lambda_k) = \nonumber\\
& \left( \sum_{j=1}^K \bm H_k^H \bm W_j \bm U_j \bm W_j^H \bm H_k \!+\! \bm G_k^H \bm A_0^{-1} \bm G_k \!+\! \lambda_k \bm I_{T_k} \!\right)^{\!\!-1} \!\!\!\bm H_k^H \bm W_k \bm U_k.
\end{align}
If matrix $\sum_{j=1}^K \bm H_k^H \bm W_j \bm U_j \bm W_j^H \bm H_k + \bm G_k^H \bm A_0^{-1} \bm G_k$ is invertible and ${\text {tr}} \left[ \bm V_k (0) \bm V_k (0)^H \right] \leq P_k$, we have $\bm V_k^* = \bm V_k (0)$.
Otherwise, $\lambda_k > 0$ and ${\text {tr}}(\bm V_k \bm V_k^H) = P_k$. 
Let $\bm \varGamma_k \bm \varLambda_k \bm \varGamma_k^H$ denote the eigendecomposition of matrix $\sum_{j=1}^K \bm H_k^H \bm W_j \bm U_j \bm W_j^H \bm H_k + \bm G_k^H \bm A_0^{-1} \bm G_k$ and $\bm \varPhi_k = \bm \varGamma_k^H \bm H_k^H \bm W_k \bm U_k \bm U_k^H \bm W_k^H \bm H_k \bm \varGamma_k $.
Then,
\begin{align}\label{trace_VV}
{\text {tr}}(\bm V_k \bm V_k^H) & = {\text {tr}} \left[ \left( \bm \varLambda_k + \lambda_k \bm I_{T_k} \right)^{-2} \bm \varPhi_k \right]\nonumber\\
& = \sum_{j=1}^{T_k} \frac{[\bm \varPhi_k]_{jj}}{ \left( [\bm \varLambda_k]_{jj} + \lambda_k \right)^2 }\nonumber\\
& = P_k.
\end{align}
It is obvious that ${\text {tr}}(\bm V_k \bm V_k^H)$ in (\ref{trace_VV}) decreases monotonically with $\lambda_k$.
Hence, the optimal $\lambda_k^*$ can be found using bisection searching method.
The optimal $\bm V_k^*$ can then be obtained from (\ref{V_k}).
The alternative algorithm for solving problem (\ref{problem1}) is summarized in Algorithm \ref{algorithm_inde}.

Once problem (\ref{problem1}) is solved using Algorithm \ref{algorithm_inde}, a sub-optimal solution of problem (\ref{C2_max}) can be obtained by setting $\bm F_k = \bm V_k \bm V_k^H, \forall k \in \cal K$.

\begin{algorithm}[h]
	\begin{algorithmic}[1]
		\caption{Alternative algorithm for solving problem (\ref{problem1})}
		\State Initialize $\bm V_k, ~\forall~ k \in \cal K$.
		\Repeat
		\State Update $\bm W_k, ~\forall~ k \in \cal K$, based on (\ref{W_k}).
		\State Update $\bm U_k, ~\forall~ k \in \cal K$, based on (\ref{U_k}).
		\Repeat
		\State Let $\bm F_{k0} = \bm V_k \bm V_k^H, ~\forall~ k \in \cal K$.
		\State Update $\bm V_k, ~\forall~ k \in \cal K$, based on (\ref{V_k}).
		\Until convergence
		\Until convergence
		\label{algorithm_inde}
	\end{algorithmic}
\end{algorithm}

\subsection{Convergence and Complexity Analysis}
\label{convergence_complexity}

\subsubsection{Convergence Analysis}
\label{convergence_analysis}

Since Algorithm~\ref{algorithm_joint} and Algorithm~\ref{algorithm_inde} are carried out in iterative manners, it is necessary to characterize their convergence behaviors. 

From \cite{7547360} and \cite{7296696}, it is known that by successive convex approximation, Algorithm~\ref{algorithm_joint} converges to a stationary point of problem (\ref{C1_max}).
As for Algorithm~\ref{algorithm_inde}, since it is a two-layer iterative method, we need to demonstrate that the iteration in each layer converges. 
In the inner iteration, MM method is applied to update $\bm V_k$.
The iteration thus converges to a stationary point of problem (\ref{problem3}).
In the outer iteration, we refer to the WMMSE scheme proposed in \cite{shi2011iteratively}, introduce auxiliary variables $\bm W_k$ as well as $\bm U_k$, and arrive at (\ref{problem1}) such that problem (\ref{C3_max}) can be solved in an iterative manner.
It can be proven similarly as \cite[Theorem~3]{shi2011iteratively} that the outer iteration of Algorithm~\ref{algorithm_inde} converges to a stationary point $(\bm V_{\cal K}, \bm W_{\cal K}, \bm U_{\cal K})$ of problem (\ref{problem1}) and the corresponding $\bm V_{\cal K}$ is a stationary point of problem (\ref{C3_max}).
Since problem (\ref{C3_max}) is equivalent to (\ref{C2_max}), letting $\bm F_k = \bm V_k \bm V_k^H, \forall k \in \cal K$, a stationary point of the original problem (\ref{C2_max}) can then be obtained.

\subsubsection{Complexity Analysis}
\label{complexity_analysis}

To evaluate the complexity of the proposed algorithms, we count the total number of floating-point operations (FLOPs), where one FLOP represents a complex multiplication or a complex summation, express it as a function (usually a polynomial) of
the dimensions of the matrices involved, and simplify the expression
by ignoring all terms except the leading (i.e., highest order or dominant) terms \cite{boyd2004convex, hunger2005floating}.
For the sake of convenience, we assume equal number of antennas for all users, i.e., $T_k = T, \forall k \in {\cal K}$, when analyzing the complexity. 
People can also use $\max \{ T_k, \forall k \in {\cal K} \}$ instead to evaluate the complexity.

The complexity of Algorithm~\ref{algorithm_joint} mainly lies in solving problem (\ref{C1_max2}).
Since (\ref{C1_max2}) can be easily transformed to the general determinant maximization optimization problem \cite[(19)]{YANG2019100730} with $KT^2$ variables (i.e., all entries in $F_k, \forall k \in {\cal K}$), a $B$-dimensional matrix inside the determinant operation, and a $KT$-dimensional constraint space, using the results in \cite[(20)]{YANG2019100730} and \cite[Section~10]{Vandenberghe1998Determinant}, Algorithm~\ref{algorithm_joint} involves a total complexity of ${\cal O}\left( \kappa_1 \sqrt{K} K^2 T^3 \left( K^2 T^4 + B^2 \right)\right)$, where $\kappa_1$ is the number of iterations of Algorithm~\ref{algorithm_joint}.

As for Algorithm~\ref{algorithm_inde}, its complexity mainly lies in calculating $\bm W_k$, $\bm U_k$, and $\bm V_k$ in (\ref{W_k}), (\ref{U_k}), and (\ref{V_k}), which involves matrix inversion, multiplication, and summation.
Let $\kappa_2$ and $\kappa_3$ respectively denote the number of the outer and inner iterations of Algorithm~\ref{algorithm_inde}.
Since the product of a $b \times c$ dimensional matrix and a $c \times d$ dimensional matrix costs ${\cal O}\left( bcd \right)$ FLOPs, and the inversion of a $b \times b$ Hermite matrix requires ${\cal O}\left( b^3 \right)$ FLOPs \cite{boyd2004convex, he2014coordinated}, Algorithm~\ref{algorithm_inde} involves a total complexity of ${\cal O}\left( \kappa_2 \left[ K^2 (BT^2 + B^2T + T^3) + KB^3 \right.\right.$ $\left.\left. + \kappa_3 \left( K (ET^2 + E^2T) + E^3 \right) \right] \right)$.
Note that since Algorithm~\ref{algorithm_inde} includes a bisection step (finding the optimal $\lambda_k^*$ in (\ref{trace_VV})) which generally takes few iterations, we ignore this bisection step in the complexity analysis.

It is worth mentioning that the given analysis only shows how the bounds on computational complexity are related to different problem dimensions. 
In practice, the actual computational load may vary depending on the structure simplifications and used numerical solvers.


\section{Simulation Results}
\label{simulation}

\begin{figure}
	\centering
	\includegraphics[scale=0.50]{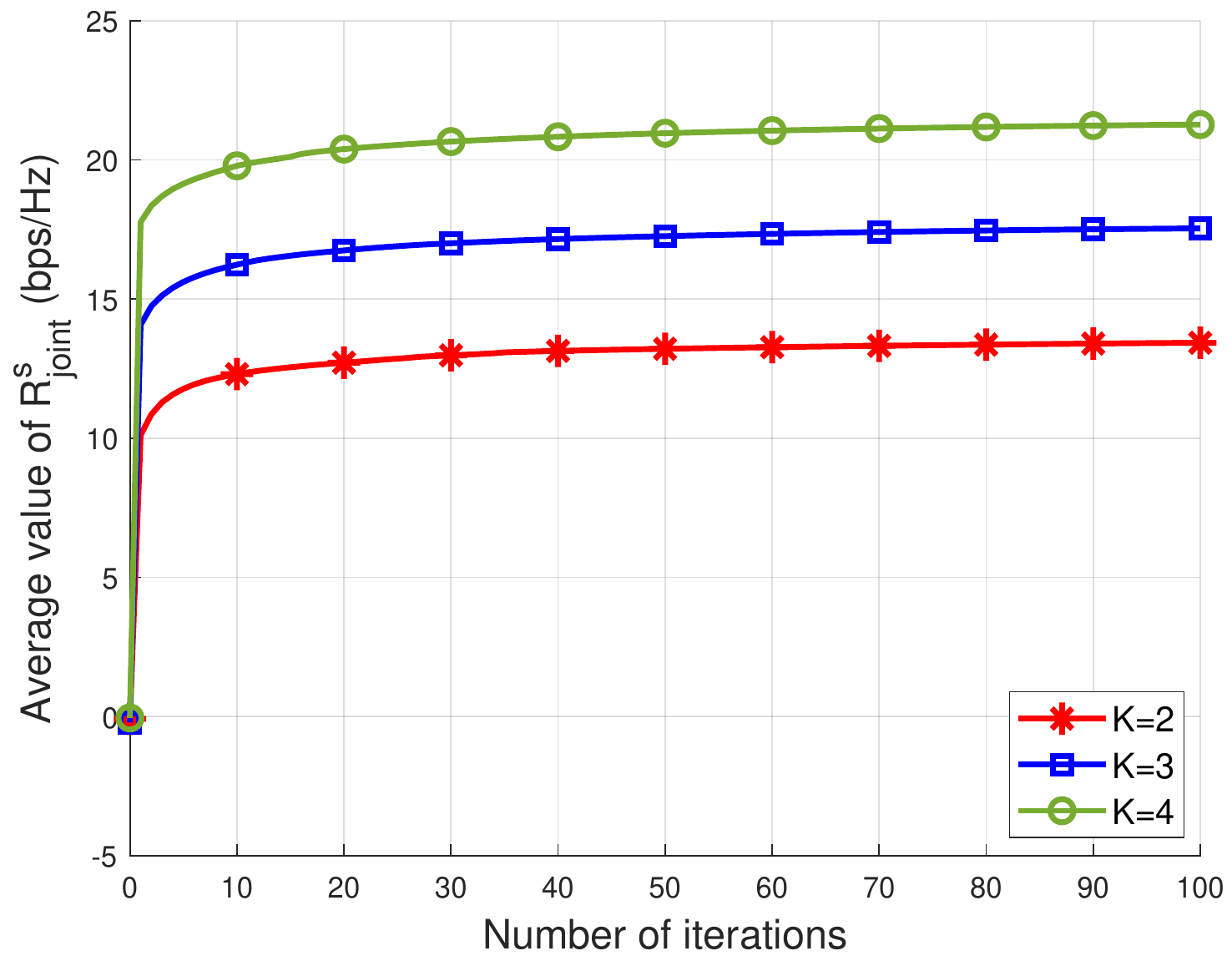}
	\caption{Convergence behaviors of Algorithm~\ref{algorithm_joint} with $T=4$, $B=E=4$, and $P=10$ dBm.}
	\label{Iterate_joint}
\end{figure}

In this section, simulation results are presented to evaluate the performance of the proposed algorithms.
We consider an isolated circular-cell network with a radius of $500$ meters.
The base station or Bob is located at the center and an Eve is evenly distributed in the cell.
All mobile users are distributed uniformly in the cell and it is assumed that no user
is closer to Bob than $20$ meters. 
For convenience, equal maximum power constraint, number of antennas at all users, and noise power at Bob and Eve, are assumed, i.e., $P_k = P$, $T_k = T, ~  \forall k\in {\cal K}$, and $\sigma_B^2 = \sigma_E^2 = \sigma^2$. 
The pathloss exponent and the standard deviation of log-normal shadowing fading are respectively set to be $3.7$ and $8$ dB \cite{access2010further}. 
The noise power is $\sigma^2 = -100$ dBm.
All simulation results are obtained by averaging over $1000$ independent channel realizations, and each channel realization is obtained by generating a random
user distribution as well as a random set of fading coefficients.

\begin{figure}
	\centering
	\includegraphics[scale=0.50]{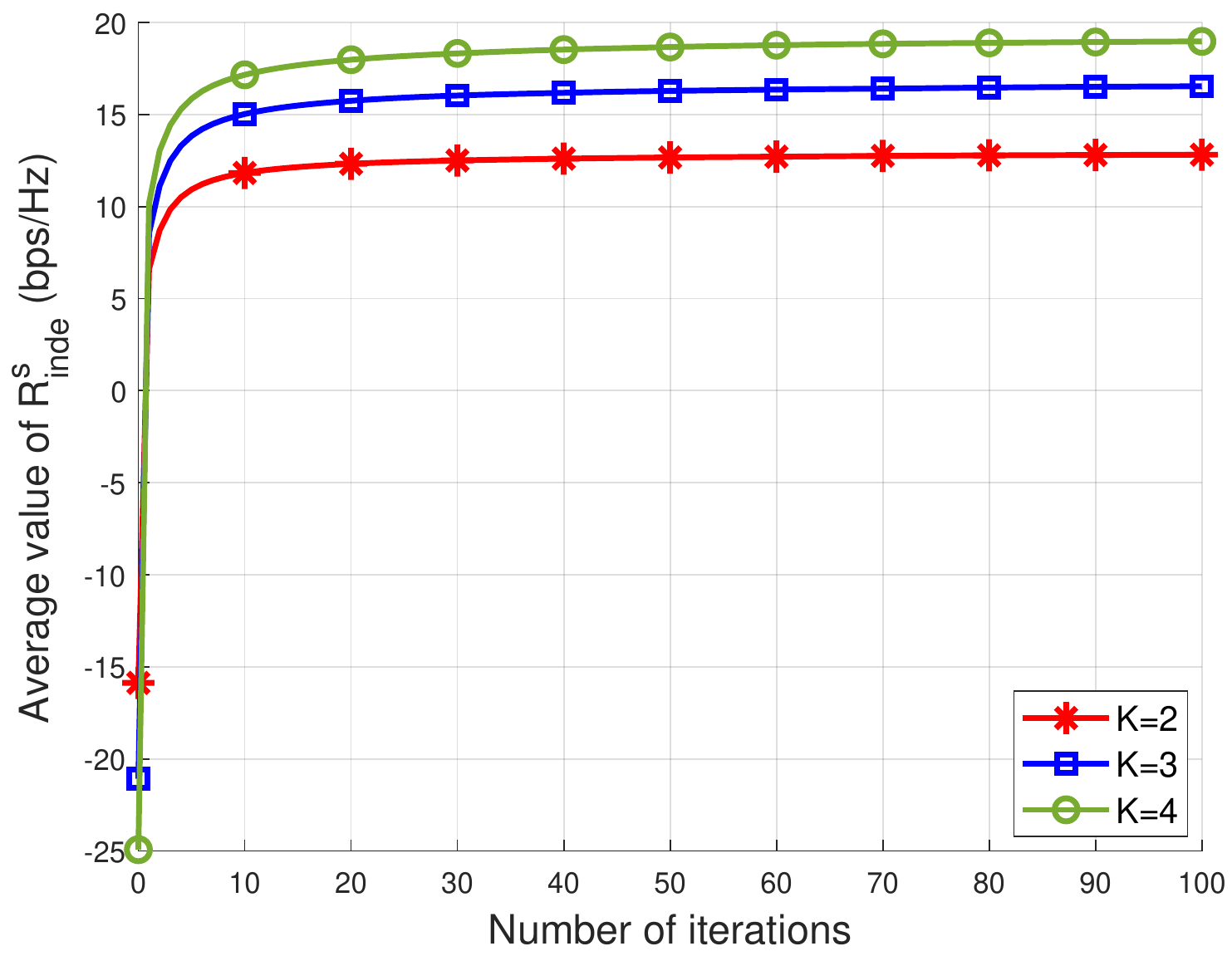}
	\caption{Convergence behaviors of Algorithm~\ref{algorithm_inde} with $T=4$, $B=E=4$, and $P=10$ dBm.}
	\label{Iterate_inde}
\end{figure}

\begin{figure}
	\centering
	\includegraphics[scale=0.50]{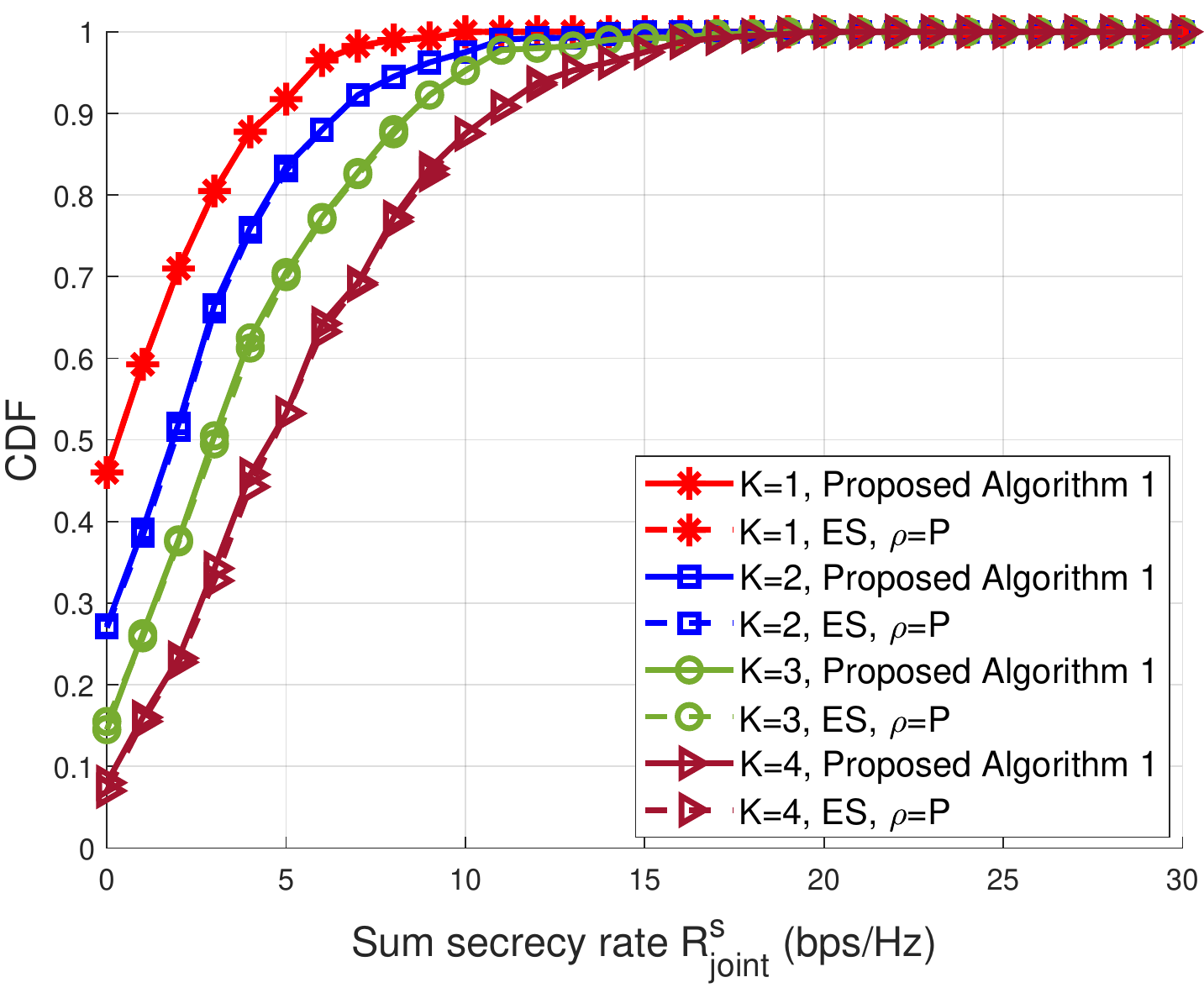}
	\caption{CDF of sum secrecy rate $R_{\text {joint}}^{\text s}$ for the proposed Algorithm~\ref{algorithm_joint} and ES with $T=1$, $B=E=4$, and $P=10$ dBm.}
	\label{CDF_VS_ES_joint}
\end{figure}
Fig.~\ref{Iterate_joint} and Fig.~\ref{Iterate_inde} illustrate the convergence behaviors of Algorithm~\ref{algorithm_joint} and the outer iteration of Algorithm~\ref{algorithm_inde}.
It can be seen from these two figures that the average values of $R_{\text {joint}}^{\text s}$ and $R_{\text {inde}}^{\text s}$ both increase greatly and monotonically during their corresponding iterative processes and converge rapidly for different configurations of $K$, which shows the significant advantages of the proposed algorithms.
Note that since $[\cdot]^+$ operation is omitted in (\ref{R_s_joint_GV}) and (\ref{R_s_inde_GV}) for convenience, the average values of $R_{\text {joint}}^{\text s}$ and $R_{\text {inde}}^{\text s}$ could be negative as shown in Fig.~\ref{Iterate_joint} and Fig.~\ref{Iterate_inde}. 
Interestingly, with the same initial covariance matrices $\{\bm F_k\}$, the initial average value of $R_{\text {joint}}^{\text s}$ is close to $0$ while that of $R_{\text {joint}}^{\text s}$ is much smaller, e.g., $-15$ bps/Hz when $K=2$.
This is because Bob applies joint decoding in Fig.~\ref{Iterate_joint} while uses independent decoding in Fig.~\ref{Iterate_inde}.

\begin{figure}
	\centering
	\includegraphics[scale=0.50]{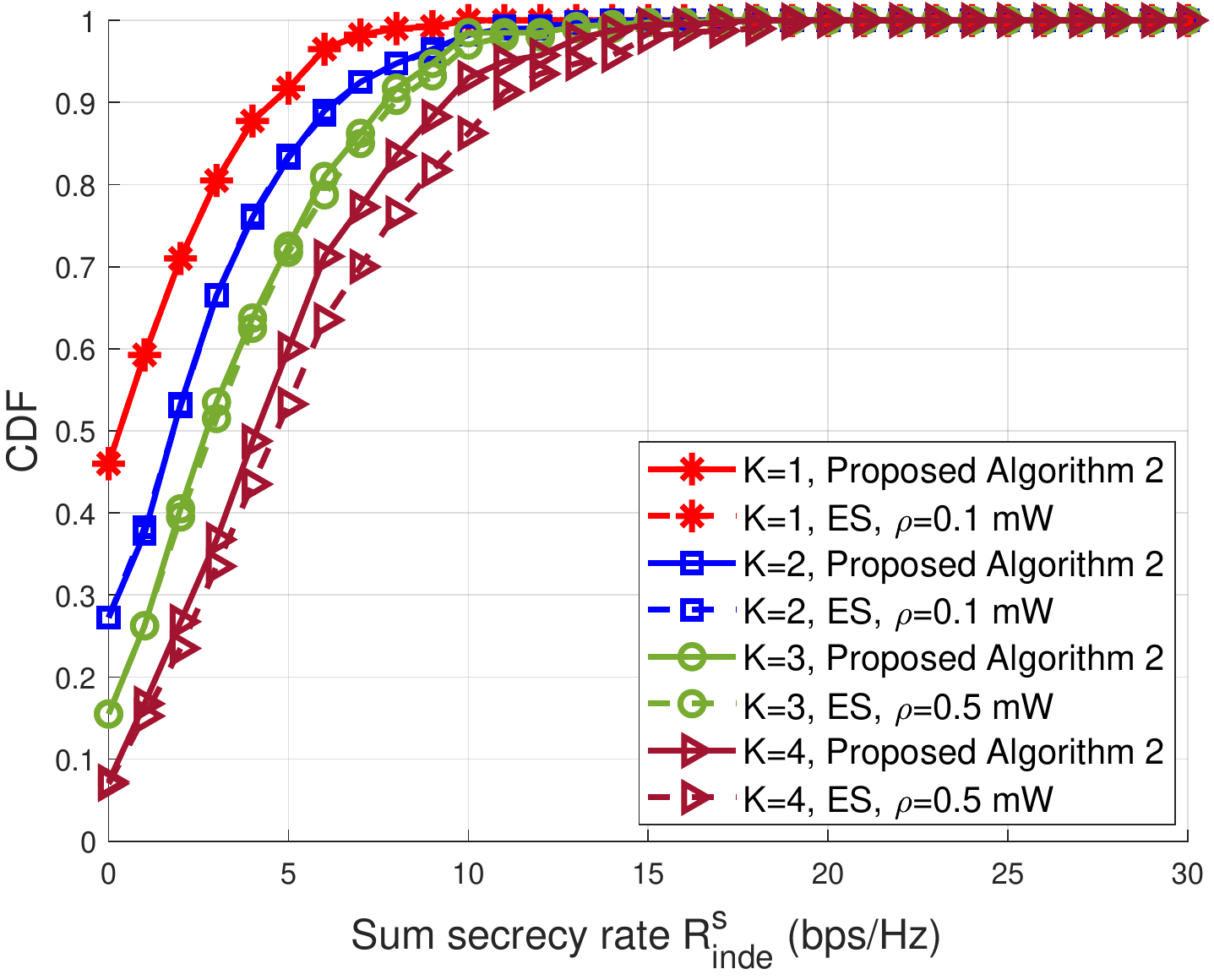}
	\caption{CDF of sum secrecy rate $R_{\text {inde}}^{\text s}$ for the proposed Algorithm~\ref{algorithm_inde} and ES with $T=1$, $B=E=4$, and $P=10$ dBm.}
	\label{CDF_VS_ES_inde}
\end{figure}

\begin{figure}
	\centering
	\includegraphics[scale=0.50]{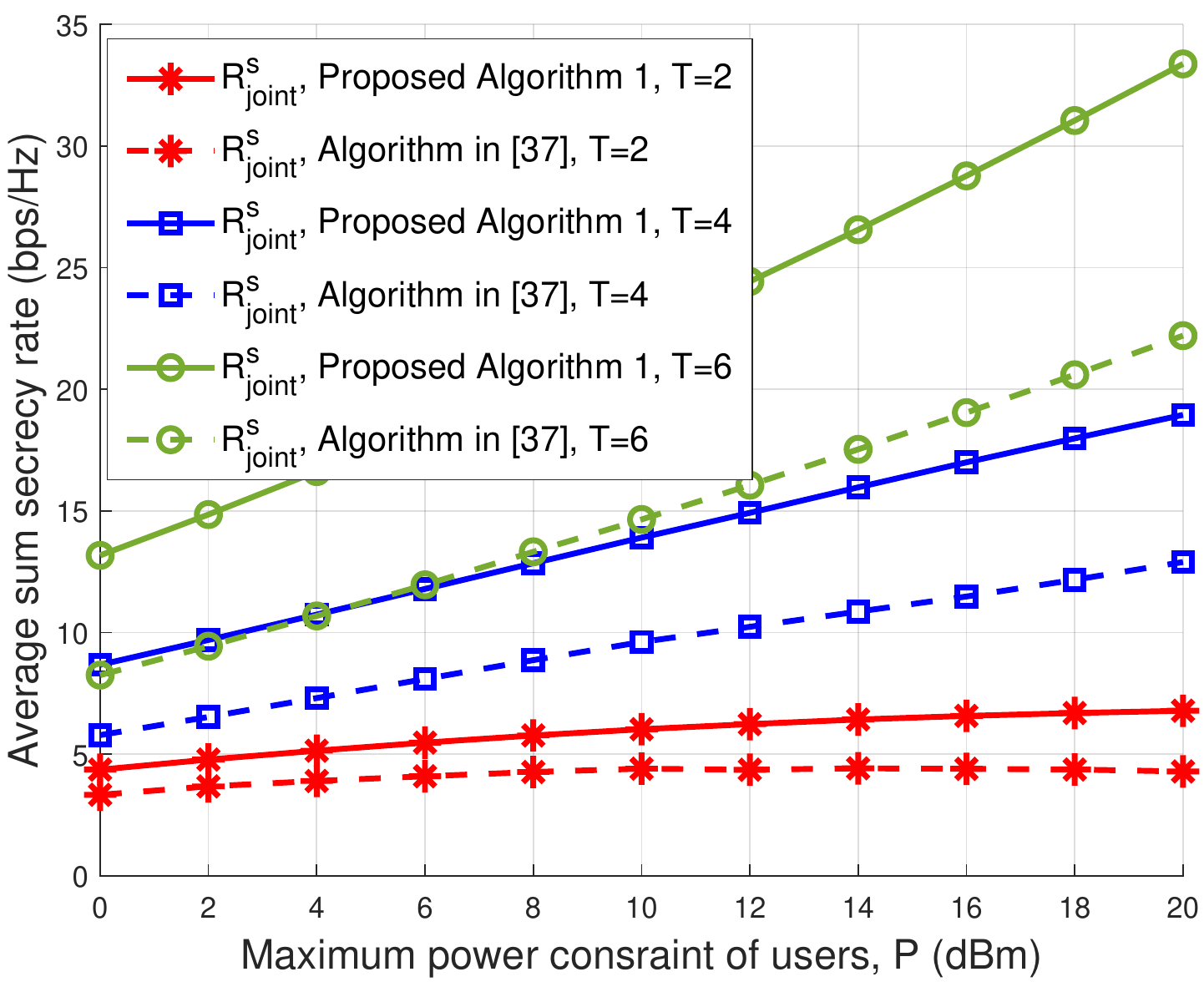}
	\caption{Average sum secrecy rate $R_{\text {joint}}^{\text s}$ versus the maximum power constraint of users with $K=2$ and $B=E=4$.}
	\label{SSR_VS_P}
\end{figure}
In Fig.~\ref{CDF_VS_ES_joint} and Fig.~\ref{CDF_VS_ES_inde}, the proposed Algorithm~\ref{algorithm_joint} and Algorithm~\ref{algorithm_inde} are compared with the ES method. 
In particular, we depict the cumulative distribution function (CDF) of $R_{\text {joint}}^{\text s}$ and $R_{\text {inde}}^{\text s}$ obtained by the proposed algorithms and the ES method.
Due to the high computational complexity in implementing ES, we consider the simple SIMO case with $T=1$.
In this case, both (\ref{C1_max}) and (\ref{C2_max}) reduce to the power control problem.
For example, as shown in Subsection~\ref{SSR_max_joint}, problem (\ref{C1_max}) can be transformed to (\ref{C1_max_SIMO}) in this case.
We thus implement the ES method by varying the transmit power of each user from $0$ to $P$ (mW) in steps of $\rho$ (mW).
Obviously, it is known from Lemma~\ref{joing_SIMO} that the ES step for solving problem (\ref{C1_max}) or (\ref{C1_max_SIMO}) should be $\rho = P$, i.e., we check power $0$ and $P$ for each user.
As for the ES method for solving problem (\ref{C2_max}), we set $\rho = 0.1$ mW for the cases with $K=1$ and $K=2$.
Since the computational complexity of ES is proportional to $(P/\rho + 1)^K$, it becomes quite time-consuming when $\rho$ is small and $K$ is large.
We thus set $\rho = 0.5$ mW for the cases with $K=3$ and $K=4$.
From Fig. \ref{CDF_VS_ES_joint} and Fig.~\ref{CDF_VS_ES_inde}, it can be found that for all considered configurations, the CDF curves obtained by Algorithm~\ref{algorithm_joint} almost coincide with those obtained by ES, and the performance gap between Algorithm~\ref{algorithm_inde} and the ES method is also limited, indicating that the proposed algorithms perform well in solving the sum secrecy rate maximization problems.
Note that when $T>1$, the ES method will become computationally intractable since its complexity increases exponentially with $K$ and $T$.
In contrast, as analyzed in Subsection~\ref{convergence_complexity}, the proposed algorithms involve much less computational complexity.

\begin{figure}
	\centering
	\includegraphics[scale=0.50]{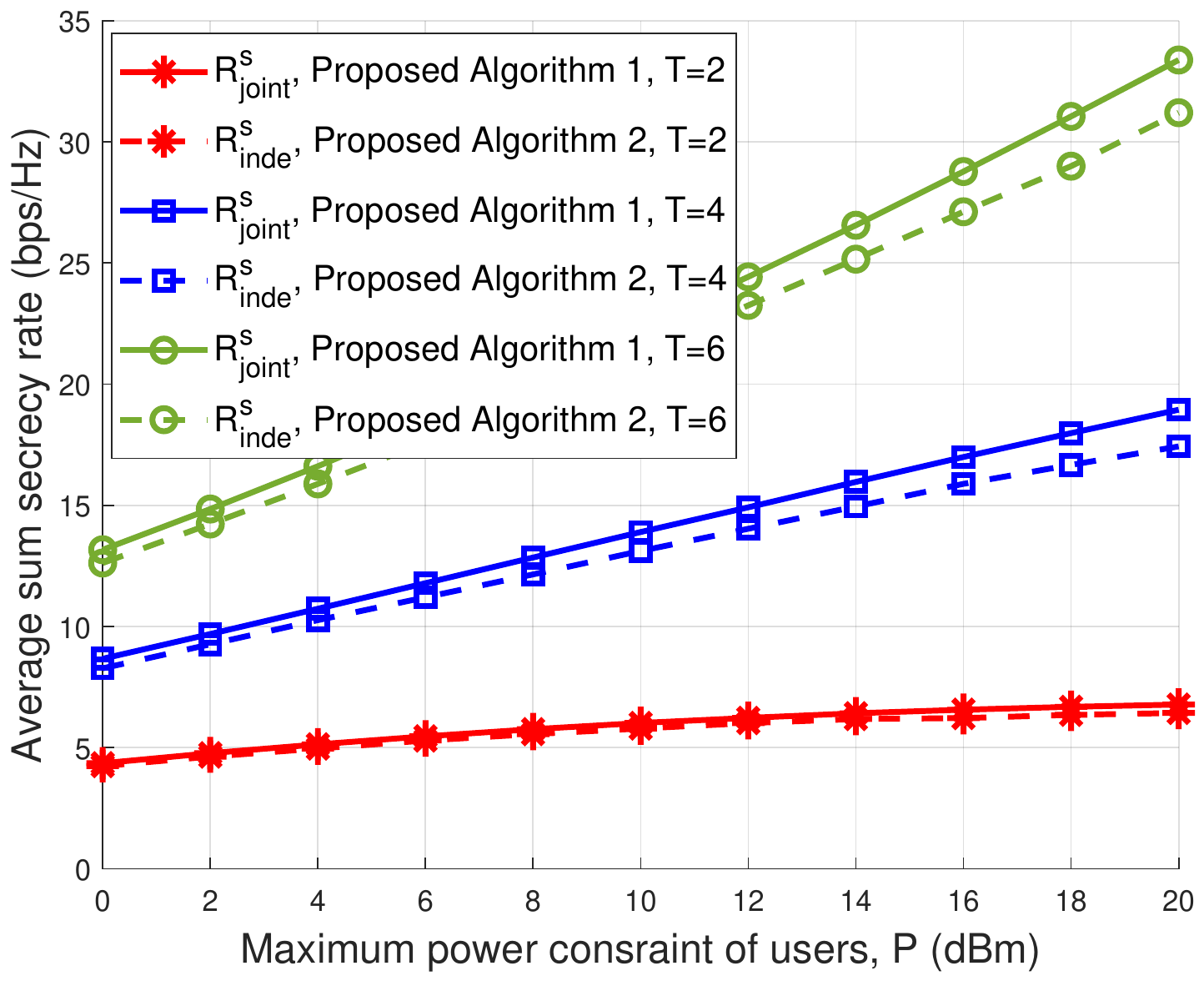}
	\caption{Average sum secrecy rate $R_{\text {joint}}^{\text s}$ and $R_{\text {inde}}^{\text s}$ versus the maximum power constraint of users with $K=2$ and $B=E=4$.}
	\label{C1C2_VS_P}
\end{figure}

\begin{figure}
	\centering
	\includegraphics[scale=0.50]{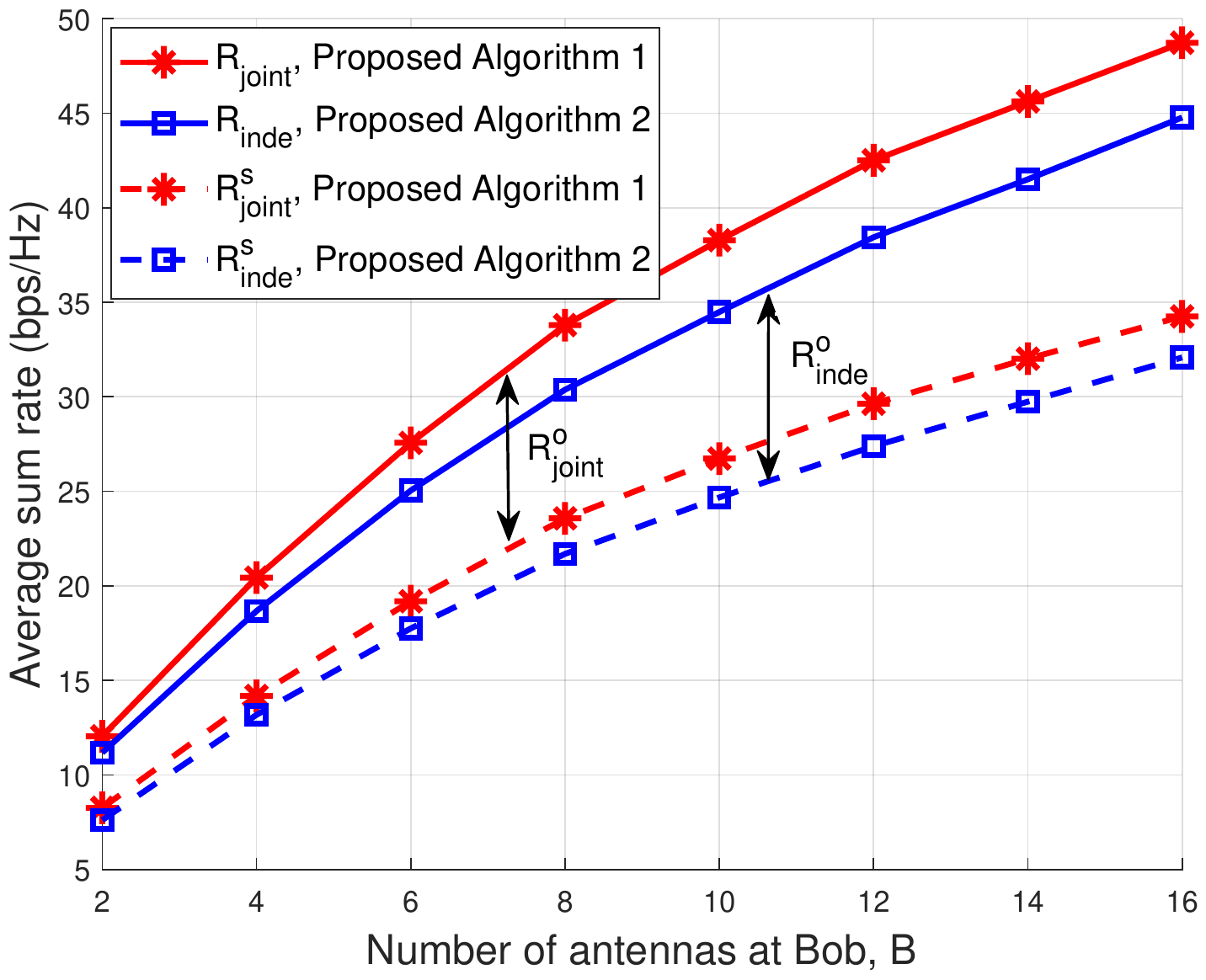}
	\caption{Average sum rate versus the number of antennas at Bob with $K=2$, $T=4$, $E=4$, and $P=10$ dBm.}
	\label{RC_VS_B}
\end{figure}
In Fig.~\ref{SSR_VS_P} and Fig.~\ref{C1C2_VS_P}, the effect of the maximum transmit power $P$ is investigated under different configurations of $T$.
For comparison, the average value of $R_{\text {joint}}^{\text s}$ obtained by \cite{lee2017precoder} is also depicted in Fig.~\ref{SSR_VS_P}.
Several observations can be made from these two figures.
First, as expected, both $R_{\text {joint}}^{\text s}$ and $R_{\text {inde}}^{\text s}$ increase with $P$ for all configurations of $T$.
Second, the average value of $R_{\text {joint}}^{\text s}$ obtained by \cite{lee2017precoder} is much smaller than that obtained by the proposed Algorithm~\ref{algorithm_joint} and the gap between them wides greatly as $P$ or $T$ increases.
As explained in Subsection~\ref{SSR_max_joint}, this is because the problem considered in \cite{lee2017precoder} is a special case of problem (\ref{C1_max}).
In addition, it can be found from Fig.~\ref{C1C2_VS_P} that the gap between $R_{\text {joint}}^{\text s}$ and $R_{\text {inde}}^{\text s}$ is small for different values of $P$ and $T$.
It shows that by using the proposed Algorithm~\ref{algorithm_inde}, a `good' sum secrecy rate $R_{\text {inde}}^{\text s}$ can be achieved even when Bob independently decodes the messages of all users.

\begin{figure}
	\centering
	\includegraphics[scale=0.50]{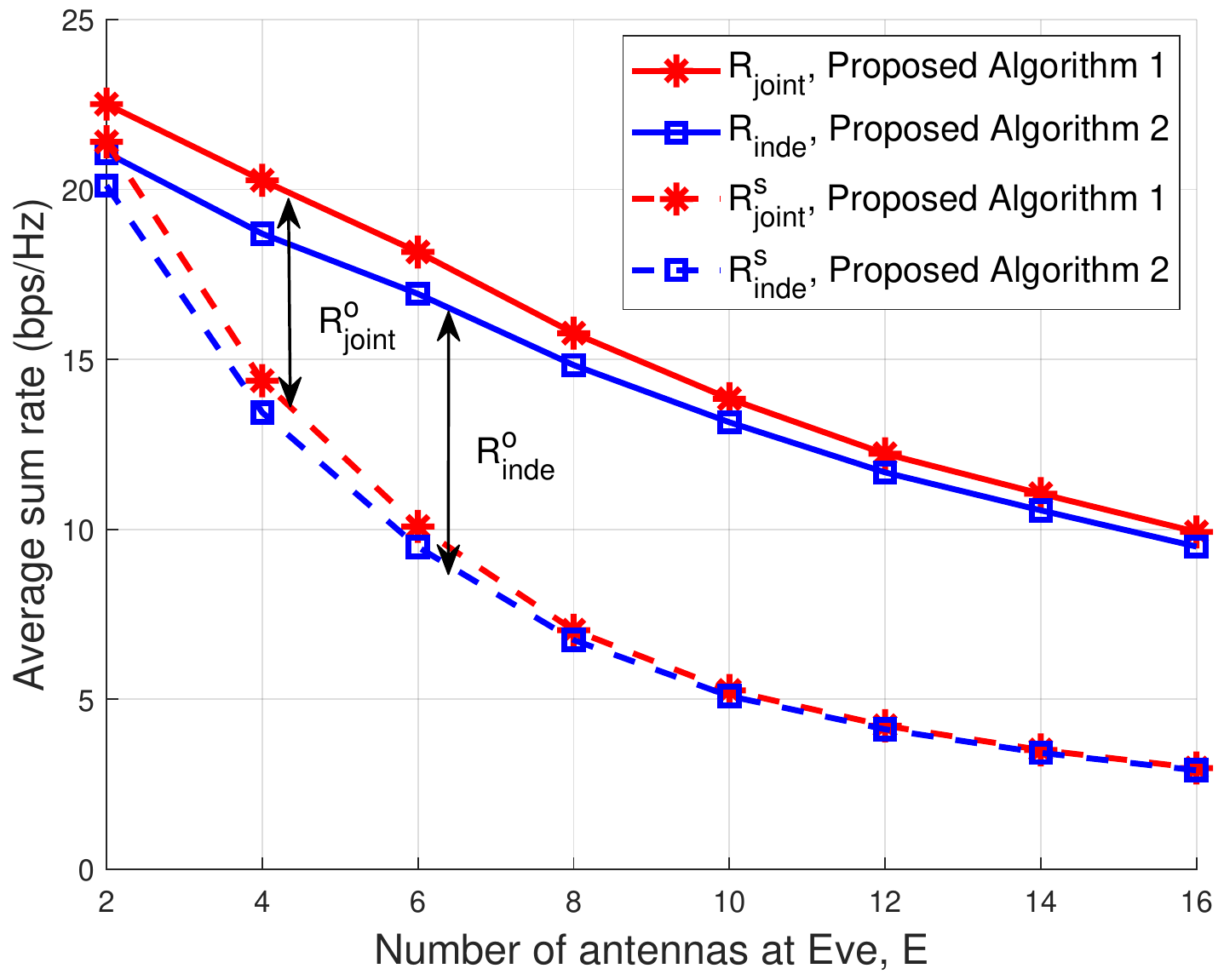}
	\caption{Average sum rate versus the number of antennas at Eve with $K=2$, $T=4$, $B=4$, and $P=10$ dBm.}
	\label{RC_VS_E}
\end{figure}

In Fig.~\ref{RC_VS_B} and Fig.~\ref{RC_VS_E}, we investigate the effect of the antenna numbers at Bob and Eve, respectively.
Besides the sum secrecy rate $R_{\text {joint}}^{\text s}$ and $R_{\text {inde}}^{\text s}$, which are the main metrics we aim to maximize in (\ref{C1_max}) and (\ref{C2_max}), we also depict the sum rate $R_{\text {joint}}$ and $R_{\text {inde}}$, and sum open rate $R_{\text {joint}}^{\text o}$ and $R_{\text {inde}}^{\text o}$ in Fig.~\ref{RC_VS_B} and Fig.~\ref{RC_VS_E}.
Since we focus on maximizing $R_{\text {joint}}^{\text s}$ and $R_{\text {inde}}^{\text s}$, as expected, they increase with $B$ and decrease with $E$.
It can also be found from these two figures that the spectrum efficiency of the considered network can be greatly improved by transmitting open messages at rate $R_{\text {joint}}^{\text o}$ or $R_{\text {inde}}^{\text o}$ simultaneously with the secret messages.


\section{Conclusions}
\label{section6}

In this paper, we first studied the capacity region of a DM MAC-WT channel where, besides confidential messages, the users have also open messages to transmit.
By using random coding, we gave achievable rate regions where the error probability of all messages at the intended receiver and the information leakage of the confidential messages to the eavesdropper vanish as the block length increases to infinity. 
We then extended the results in the DM case to a GV MAC-WT channel and also corrected the result in \cite{tekin2008general} that studied the SISO Gaussian MAC-WT case, but where the provided achievable region is actually not generally achievable.
Based on the information theoretic results, we further maximized the sum secrecy rate of the GV MAC-WT channel.
Simulation results showed that secret communication could be greatly enhanced by the proposed algorithms and the spectrum efficiency of the system could be improved since open messages were transmitted simultaneously with the secret messages.

\appendices

\section{Proof of Theorem~\ref{theorem1}}
\label{Prove_theorem1}

\begin{figure*}[ht]
	\centering
	\includegraphics[scale=0.95]{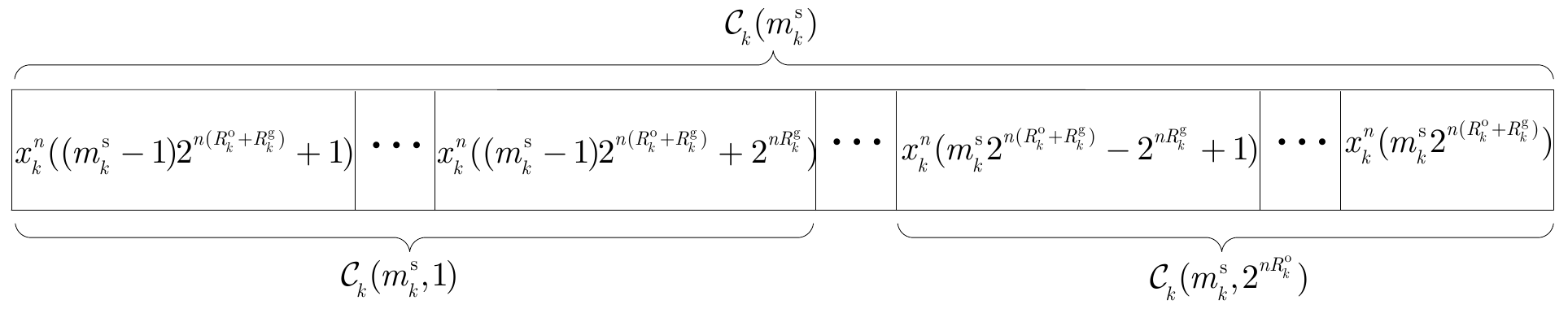}
	\caption{A division of subcodebook ${\cal C}_k(m_k^{\text s})$.}
	\label{Subcodebook}
\end{figure*}

In this appendix, we show that there exists a $\left( 2^{n R_1^{\text s}}, 2^{n R_1^{\text o}}, 2^{n R_2^{\text s}}, 2^{n R_2^{\text o}}, n \right)$ code such that any rate tuple inside region ${\mathscr R} (X_1, X_2)$, i.e., any $(R_1^{\text s}, R_1^{\text o}, R_2^{\text s}, R_2^{\text o})$ satisfying
\begin{equation}\label{rate_region1}
\left\{\!\!\!
\begin{array}{ll}
\sum\limits_{k \in {\cal S}} (R_k^{\text s} + R_k^{\text o}) < I(X_{\cal S}; Y| X_{\bar {\cal S}}) - \epsilon, \forall \cal S \subseteq \cal K, \\
\sum\limits_{k \in {\cal S}} R_k^{\text s} \!<\! I(X_{\cal S}; Y| X_{\bar {\cal S}}) \!-\! I(X_{\cal S}; Z) \!-\! \epsilon, \!\forall \cal S \subseteq \cal K,\\
R_1^{\text s} \!+\! R_2^{\text s} \!+\! R_k^{\text o} \!<\! I(X_{\cal K}; Y) \!-\! I(X_{\bar k}; Z) \!-\! \epsilon, \forall k \in \cal K,
\end{array} \right.\!\!
\end{equation}
is achievable, where $\epsilon$ is an arbitrarily small positive number. 
This, together with the standard time-sharing over coding strategies, suffices to prove the theorem. 
Note that we assume $I(X_{\cal S}; Y| X_{\bar {\cal S}}) > I(X_{\cal S}; Z), ~\forall~ \cal S \subseteq \cal K$ in (\ref{rate_region1}) for convenience.
This naturally ensures $I(X_{\cal K}; Y) > I(X_{\bar k}; Z), \forall k \in \cal K$.
If this assumption is not true, we explain after Lemma~\ref{lemma0} that Theorem~\ref{theorem1} can still be proven by simply modifying the proof steps provided in this appendix.

\begin{lemma}\label{lemma0}
	For any rate tuple $(R_1^{\text s}, R_1^{\text o}, R_2^{\text s}, R_2^{\text o})$ satisfying (\ref{rate_region1}), there exists a rate pair $(R_1^{\text g}, R_2^{\text g})$ such that
	\begin{equation}\label{rate_region2}
	\left\{\!\!\!
	\begin{array}{ll}
	R_k^{\text g} \geq 0, ~\forall~ k \in \cal K, \\
	\sum\limits_{k \in {\cal S}} (R_k^{\text s} \!+\! R_k^{\text o} \!+\! R_k^{\text g}) \!<\! I(X_{\cal S}; Y| X_{\bar {\cal S}}) \!-\! \epsilon, \forall \cal S \subseteq \cal K, \\
	\sum\limits_{k \in {\cal S}} (R_k^{\text o} + R_k^{\text g}) \geq I(X_{\cal S}; Z), \forall \cal S \subseteq \cal K.
	\end{array} \right.
	\end{equation}
\end{lemma}
\itshape \textbf{Proof:}  \upshape
By eliminating $R_1^{\text g}$ and $R_2^{\text g}$ in (\ref{rate_region2}) using the Fourier-Motzkin procedure \cite[Appendix D]{el2011network}, it can be shown that (\ref{rate_region1}) is the projection of (\ref{rate_region2}) onto the hyperplane $\{ R_1^{\text g} = 0, R_2^{\text g} = 0\}$. 
Lemma \ref{lemma0} can thus be proven.
Due to space limitation, the detailed elimination procedure is omitted.
\hfill $\Box$

In Lemma~\ref{lemma0}, we introduce a `garbage' message to each user to interfere with the decoding of Eve, i.e., besides the secret and open messages, each user also has to transmit a `garbage' message at rate $R_k^{\text g}$ though it is not necessary for Bob.
The rate of `garbage' messages $R_k^{\text g}, \forall k \in \cal K$ satisfies (\ref{rate_region2}), which is the key point for proving Theorem~\ref{theorem1}.
In particular, the second inequation of (\ref{rate_region2}) ensures that the messages (even after adding `garbage' messages) of all users can be perfectly decoded at Bob, and with the third inequation in (\ref{rate_region2}), the secret messages of all users can be perfectly protected from Eve.

Now let's recall the assumption $I(X_{\cal S}; Y| X_{\bar {\cal S}}) > I(X_{\cal S}; Z), ~\forall~ \cal S \subseteq \cal K$ made in (\ref{rate_region1}).
If, for example, $I(X_1; Y| X_2) \leq I(X_1; Z)$, we have $R_1^{\text s} = 0$.
In this case, user~$1$ cannot transmit any secret message.
Then, from the proof process in this appendix it is known that there is no need for user~$1$ to transmit the `garbage' message, i.e., $R_1^{\text g} = 0$.
It can be readily verified by using the Fourier-Motzkin procedure that for any rate tuple $(0, R_1^{\text o}, R_2^{\text s}, R_2^{\text o})$ satisfying (\ref{rate_region1}), there exists a `garbage' message rate $R_2^{\text g}$ for user~$2$ such that
\begin{equation}
\left\{\!\!\!
\begin{array}{ll}
R_2^{\text g} \geq 0, \\
R_1^{\text o} < I(X_1; Y| X_2) - \epsilon, \\
R_2^{\text s} + R_2^{\text o} + R_2^{\text g} < I(X_2; Y| X_1) - \epsilon, \\
R_1^{\text o} + R_2^{\text s} + R_2^{\text o} + R_2^{\text g} < I(X_1, X_2; Y) - \epsilon, \\
R_2^{\text o} + R_2^{\text g} \geq I(X_2; Z).
\end{array} \right.
\end{equation}
Then, Theorem~\ref{theorem1} can be proven by simply modifying the proof steps in this appendix.
For any other subsets $\cal S \subseteq \cal K$ which gives $I(X_{\cal S}; Y| X_{\bar {\cal S}}) \leq I(X_{\cal S}; Z)$, we may take similar measures to prove Theorem~\ref{theorem1}.

In the following, we prove that any rate tuple $(R_1^{\text s}, R_1^{\text o}, R_2^{\text s}, R_2^{\text o})$ satisfying (\ref{rate_region1}) is achievable.
Specifically, we first provide a random coding scheme, and then show that all users can communicate with Bob with arbitrarily small probability of error, while the confidential information leaked to Eve tends to zero.

\subsection{Coding Scheme}
\label{Prove_theorem1_A}

For a given rate tuple $(R_1^{\text s}, R_1^{\text o}, R_2^{\text s}, R_2^{\text o})$ inside region ${\mathscr R} (X_1, X_2)$, choose a rate pair $(R_1^{\text g}, R_2^{\text g})$ satisfying (\ref{rate_region2}).
Without loss of generality (w.l.o.g.), assume that $2^{n R_k^{\text s}}$, $2^{n (R_k^{\text o} + R_k^{\text g})}$ and $2^{n R_k^{\text g}}, \forall k \in {\cal K}$ are integers.
Denote
\begin{align}\label{L2}
& {\cal L}_{k,m_k^{\text s}} = \left[ (m_k^{\text s}-1) 2^{n (R_k^{\text o} + R_k^{\text g})} + 1 : m_k^{\text s} 2^{n (R_k^{\text o} + R_k^{\text g})} \right],\nonumber\\
& \quad\quad\quad \forall k \in {\cal K}, m_k^{\text s} \in {\cal M}_k^{\text s},\nonumber\\
& {\cal L}_k = \left\{ {\cal L}_{k,m_k^{\text s}}, \forall m_k^{\text s} \in {\cal M}_k^{\text s} \right\}\nonumber\\
& \quad\,= \left[ 1:2^{n (R_k^{\text s} + R_k^{\text o} + R_k^{\text g})} \right], ~\forall~ k \in {\cal K}.
\end{align}
Then, a coding scheme is provided below.

\begin{figure}
	\centering
	\includegraphics[scale=0.70]{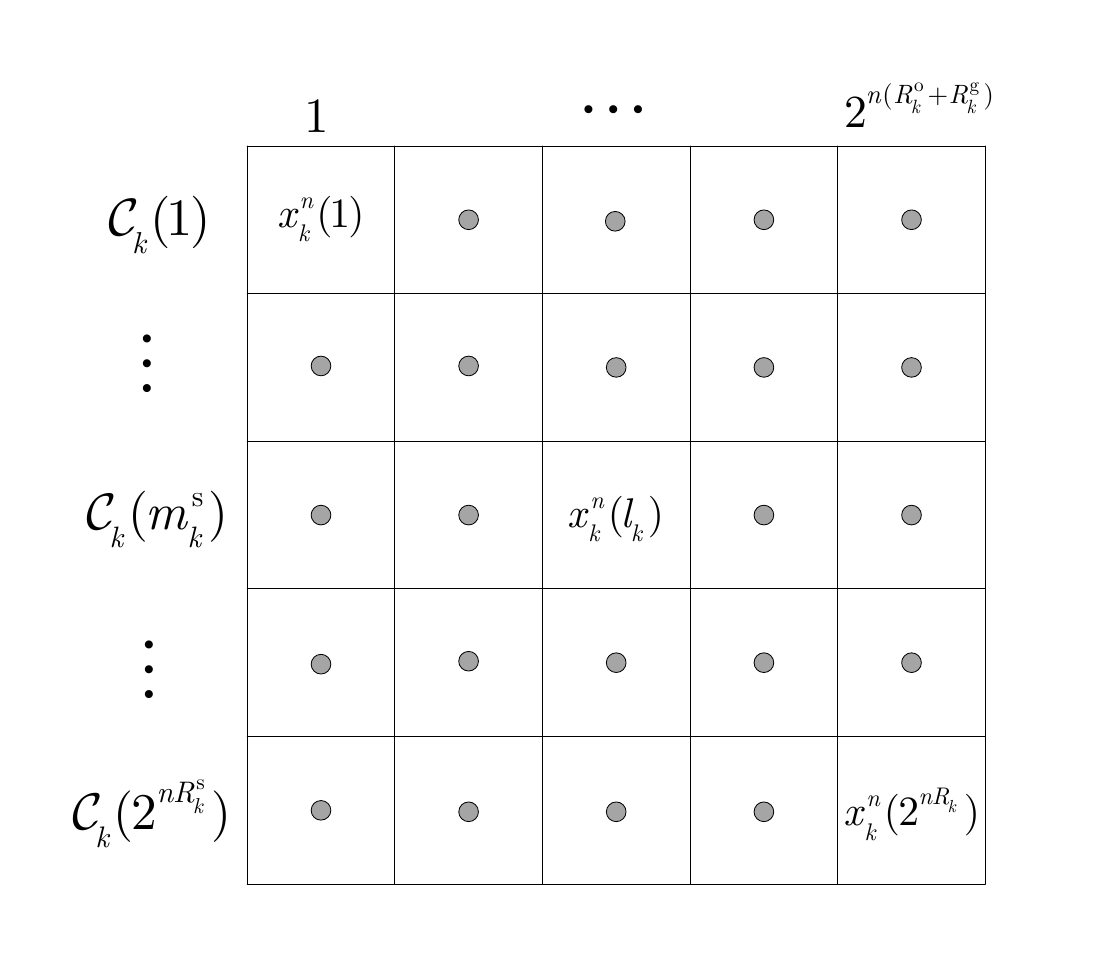}
	\caption{Codebook ${\cal C}_k$ of user $k$, where $R_k \triangleq R_k^{\text s} + R_k^{\text o} + R_k^{\text g}$.}
	\label{Codebook}
\end{figure}

{\textbf {Codebook generation.}} 
For each message pair $(m_k^{\text s}, m_k^{\text o}) \in {\cal M}_k^{\text s} \times {\cal M}_k^{\text o}$ of user $k$, generate a subcodebook ${\cal C}_k (m_k^{\text s})$ by randomly and independently generating $2^{n (R_k^{\text o} + R_k^{\text g})}$ sequences $x_k^n(l_k)$ according to $\prod_{i=1}^n p (x_{ki})$, where $l_k \in {\cal L}_{k,m_k^{\text s}}$. 
As shown in Fig. \ref{Codebook}, these subcodebooks constitute the codebook of user $k$, i.e., ${\cal C}_k = \left\{ {\cal C}_k (m_k^{\text s}), \forall m_k^{\text s} \in {\cal M}_k^{\text s} \right\}$.
The codebooks of all users, i.e., ${\cal C}_k, \forall k \in {\cal K}$, are then revealed to all transmitters and receivers, including the eavesdropper.

{\textbf {Encoding.}} 
Since $R_k^{\text g} \geq 0, \forall k \in \cal K$, as depicted in Fig. \ref{Subcodebook}, evenly divide each subcodebook 
${\cal C}_k(m_k^{\text s})$ into $2^{n R_k^{\text o}}$ subsets ${\cal C}_k(m_k^{\text s},m_k^{\text o})$ of size $2^{n R_k^{\text g}}$ codewords each, for any $m_k^{\text o} \in {\cal M}_k^{\text o}$.
To send message pair $(m_k^{\text s}, m_k^{\text o}) \in {\cal M}_k^{\text s} \times {\cal M}_k^{\text o}$, encoder $k$ uniformly chooses a codeword (with index $l_k$) from ${\cal C}_k (m_k^{\text s},m_k^{\text o})$ and then transmits $x_k^n (l_k)$.

{\textbf {Decoding.}} 
The decoder at the legitimate receiver uses joint typicality decoding to find an estimate of the messages and declares that $({\hat m}_1^{\text s}, {\hat m}_1^{\text o}, {\hat m}_2^{\text s}, {\hat m}_2^{\text o})$ is sent if it is the unique message tuple such that $(x_1^n (l_1), x_2^n (l_2), y^n) \in {\cal T}_\epsilon^{(n)}(X_1,X_2,Y)$, for some $l_1$ and $l_2$ such that $x_k^n (l_k) \in {\cal C}_k ({\hat m}_k^{\text s},{\hat m}_k^{\text o})$,  for $ k = 1,2$. 

\subsection{Analysis of the Probability of Error}

Since $\sum_{k \in {\cal S}} (R_k^{\text s} + R_k^{\text o} + R_k^{\text g}) < I(X_{\cal S}; Y| X_{\bar {\cal S}}) - \epsilon, \forall \cal S \subseteq \cal K$, it can be proven by using law of large numbers (LLN) and the packing lemma that the probability of error averaged over the random codebook and encoding tends to zero as $n \rightarrow \infty$.
The proof follows exactly the same steps used in \cite[Subsection 4.5.1]{el2011network}.
Hence, $\lim_{n \rightarrow \infty} P_{\text e} \leq \delta$.

\subsection{Analysis of the Information Leakage Rate}
\label{A-C}

For a given codebook ${\cal C}_k$, the secret message $M_k^{\text s}$ is a function of the codeword index $L_k$.
Hence,
\begin{align}\label{IM1M2Z}
& I(M_1^{\text s}, M_2^{\text s}; Z^n) \nonumber\\
= & H(M_1^{\text s}) + H(M_2^{\text s}) - H(M_1^{\text s}, M_2^{\text s}| Z^n)\nonumber\\
= & n R_1^{\text s} + n R_2^{\text s} - H(L_1, L_2| Z^n) + H(L_1, L_2| M_1^{\text s}, M_2^{\text s}, Z^n).
\end{align}
In order to measure the information leakage rate (\ref{IM1M2Z}), we first transform $H(L_1, L_2| Z^n)$ as follows
\begin{align}\label{HL1L2Z}
& H(L_1, L_2| Z^n) \nonumber\\
= & H(L_1, L_2) - I(L_1, L_2; Z^n)\nonumber\\
\overset{(a)}{=} & H(L_1) + H(L_2) - I(L_1, L_2, X_1^n, X_2^n; Z^n)\nonumber\\
\overset{(b)}{=} & n (R_1^{\text s} + R_1^{\text o} + R_1^{\text g} + R_2^{\text s} + R_2^{\text o} + R_2^{\text g}) - I(X_1^n, X_2^n; Z^n)\nonumber\\
\overset{(c)}{=} & n (R_1^{\text s} + R_1^{\text o} + R_1^{\text g} + R_2^{\text s} + R_2^{\text o} + R_2^{\text g}) - n I(X_1, X_2; Z),
\end{align}
where $(a)$ holds since $X_1^n$ and $X_2^n$ are respectively functions of indexes $L_1$ and $L_2$, $(b)$ holds since $(L_1, L_2) \rightarrow (X_1^n, X_2^n) \rightarrow Z^n$ forms a Markov chain, and $(c)$ follows since $p(x_1^n, x_2^n, z^n) = \prod_{i=1}^n p (x_{1i}, x_{2i}, z_i)$.
Then, we provide an upper bound to term $H(L_1, L_2| M_1^{\text s}, M_2^{\text s}, Z^n)$ in (\ref{IM1M2Z}) in the following theorem.
\begin{theorem}\label{lemma2}
	Using the coding scheme provided in Appendix~\ref{Prove_theorem1_A}, we have the following inequation
	\begin{align}\label{upper_bound}
	& \lim_{n \rightarrow \infty} \frac{1}{n} H(L_1, L_2| M_1^{\text s}, M_2^{\text s}, Z^n)\nonumber\\
	& \leq R_1^{\text o} + R_1^{\text g} + R_2^{\text o} + R_2^{\text g} - I(X_1, X_2; Z) + \delta.
	\end{align}
\end{theorem}

\itshape \textbf{Proof:}  \upshape
See Appendix \ref{Appendix_B}.
\hfill $\Box$

Substituting (\ref{HL1L2Z}) and (\ref{upper_bound}) into (\ref{IM1M2Z}), we have
\begin{equation}\label{I_leq_delta}
\lim_{n \rightarrow \infty} R_{{\text E}, {\cal K}} = \lim_{n \rightarrow \infty} \frac{1}{n} I(M_1^{\text s}, M_2^{\text s}; Z^n) \leq \delta.
\end{equation}

Theorem \ref{theorem1} is thus proven.
 
\section{Proof of Theorem \ref{lemma2}}
\label{Appendix_B}

For given $n$-th order product distribution on ${\cal X}^n_1 \times {\cal X}^n_2 \times {\cal Z}^n$, 
recall the definition of conditional $\epsilon$-typical sets
	\begin{align}
	&\!{\cal T}_\epsilon^{(n)} (X_1, X_2| z^n)\nonumber\\
	&\! = \left\{ (x_1^n, x_2^n)| (x_1^n, x_2^n, z^n) \in {\cal T}_\epsilon^{(n)} (X_1, X_2, Z) \right\},\label{T_conditional}\\
	& \!{\cal T}_\epsilon^{(n)}\! (X_2| x_1^n, z^n) \!=\! \left\{\! x_2^n| (x_1^n, x_2^n, z^n) \!\in\! {\cal T}_\epsilon^{(n)}\! (X_1, X_2, Z) \!\right\}\!,\!\!\! \label{T_conditional2}\\
	& \!{\cal T}_\epsilon^{(n)}\! (X_1| x_2^n, z^n) \!=\! \left\{\! x_1^n| (x_1^n, x_2^n, z^n) \!\in\! {\cal T}_\epsilon^{(n)}\! (X_1, X_2, Z) \!\right\}\!.\!\!\! \label{T_conditional3}
	\end{align}
To prove Theorem \ref{lemma2}, we bound $H(L_1, L_2| m_1^{\text s}, m_2^{\text s}, Z^n)$ for every secret message pair $(m_1^{\text s}, m_2^{\text s})$.
First, for a given received signal $z^n$ at the eavesdropper, assume that it is a typical sequence, i.e., $z^n \in {\cal T}_\epsilon^{(n)} (Z)$, and define
\begin{align}\label{D}
{\cal D} (m_1^{\text s}, m_2^{\text s}, z^n) \!=\!& \left\{\! (l_1, l_2)| (x_1^n (l_1), x_2^n (l_2)) \!\in\! {\cal T}_\epsilon^{(n)} (X_1, X_2| z^n), \right.\nonumber\\
& \left. \forall~ (l_1, l_2) \in {\cal L}_{1,m_1^{\text s}} \times {\cal L}_{2,m_2^{\text s}} \right\},
\end{align}
and
\begin{equation}\label{Q}
Q(m_1^{\text s}, m_2^{\text s}, z^n) = \left|{\cal D} (m_1^{\text s}, m_2^{\text s}, z^n)\right|.
\end{equation}
In the following theorem, we give an upper bound to the expectation and the variance of $Q(m_1^{\text s}, m_2^{\text s}, z^n)$.
\begin{theorem}\label{theorem3}
The expectation and variance of $Q(m_1^{\text s}, m_2^{\text s}, z^n)$ can be bounded as
\begin{align}
{\mathbb E} \left[ Q(m_1^{\text s}, m_2^{\text s}, z^n) \right] & \leq 2^{n (\Delta + \delta_1 (\epsilon))}, \label{EN}\\
{\text {Var}} \left[ Q(m_1^{\text s}, m_2^{\text s}, z^n) \right] & \leq 2^{n (\Delta + \delta_1 (\epsilon))} + \sum_{k \in \cal K} 2^{n (2 \Delta - \Delta_k + \delta_1 (\epsilon))},\label{VarN}
\end{align}
where $\delta_1 (\epsilon) = 5 \epsilon$ is defined in (\ref{p_1_up}), and
\begin{align}\label{delta}
& \Delta = R_1^{\text o} + R_1^{\text g} + R_2^{\text o} + R_2^{\text g} - I(X_1, X_2; Z),\nonumber\\
& \Delta_k = R_k^{\text o} + R_k^{\text g} - I(X_k; Z), ~\forall~ k \in {\cal K}.
\end{align}
\end{theorem}

\itshape \textbf{Proof:}  \upshape
	See Appendix \ref{Appendix_C}.
\hfill $\Box$

Next, define the event 
\begin{equation}
{\cal E} (m_1^{\text s}, m_2^{\text s}, z^n) = \left\{  Q(m_1^{\text s}, m_2^{\text s}, z^n) \geq 2^{n (\Delta + \delta_1 (\epsilon)) + 1} \right\}.
\end{equation}
We have
\begin{align}\label{p_event}
& P\left\{{\cal E} (m_1^{\text s}, m_2^{\text s}, z^n)\right\} \nonumber\\
= & P \left\{  Q(m_1^{\text s}, m_2^{\text s}, z^n) \geq 2^{n (\Delta + \delta_1 (\epsilon)) + 1} \right\}\nonumber\\
\leq & P \left\{  Q(m_1^{\text s}, m_2^{\text s}, z^n) \geq {\mathbb E} \left[ Q(m_1^{\text s}, m_2^{\text s}, z^n) \right] + 2^{n (\Delta + \delta_1 (\epsilon))} \right\}\nonumber\\
\leq & P \left\{ \left| Q(m_1^{\text s}, m_2^{\text s}, z^n) \!-\! {\mathbb E} \left[ Q(m_1^{\text s}, m_2^{\text s}, z^n) \right] \right| \!\geq\! 2^{n (\Delta + \delta_1 (\epsilon))} \right\}\nonumber\\
\overset{(a)}{\leq} & \frac{{\text {Var}} \left[ Q(m_1^{\text s}, m_2^{\text s}, z^n) \right]}{2^{2n (\Delta + \delta_1 (\epsilon))}}\nonumber\\
\overset{(b)}{\leq} & 2^{- n (\Delta + \delta_1 (\epsilon))} + \sum_{k \in \cal K} 2^{ - n (\Delta_k + \delta_1 (\epsilon))},
\end{align}
where step $(a)$ follows by applying the Chebyshev inequality, and $(b)$ follows by (\ref{VarN}).
Due to (\ref{rate_region2}), $\Delta \geq 0$ and $\Delta_k \geq 0, \forall k \in {\cal K}$.
Then, it is obvious that $P\left\{{\cal E} (m_1^{\text s}, m_2^{\text s}, z^n)\right\} \rightarrow 0$ as $n \rightarrow \infty$.
For any $z^n \in {\cal T}_\epsilon^{(n)} (Z)$, define indicator variable
\begin{equation}\label{indicator_E}
O (m_1^{\text s}, m_2^{\text s}, z^n) \!=\! \left\{\!\!\!
\begin{array}{ll}
1,& \!\! {\text {if}}~ {\cal E} (m_1^{\text s}, m_2^{\text s}, z^n) ~{\text {occurs}},\\
0,& \!\! {\text {otherwise}}.\\
\end{array} \right.
\end{equation} 
Then, $P \left\{ O (m_1^{\text s}, m_2^{\text s}, z^n) = 1 \right\} \rightarrow 0$ as $n \rightarrow \infty$.

Since there are $2^{n(R_k^{\text o} + R_k^{\text g})}$ codewords in each subcodebook ${\cal C}_k (m_k^{\text s}), \forall k \in \cal K$, we have
\begin{align}\label{HL1L2Zm1m21}
& H(L_1, L_2| m_1^{\text s}, m_2^{\text s}, z^n) \nonumber\\
\leq & \log (2^{n(R_1^{\text o} + R_1^{\text g} + R_2^{\text o} + R_2^{\text g})})\nonumber\\
= & n(R_1^{\text o} + R_1^{\text g} + R_2^{\text o} + R_2^{\text g}), ~\forall~ z^n \in {\cal Z}^n,
\end{align}
and 
\begin{align}\label{HL1L2Zm1m23}
& H(L_1, L_2| m_1^{\text s}, m_2^{\text s}, Z^n) \nonumber\\
= & \sum_{z^n \in {\cal Z}^n} p(z^n) H(L_1, L_2| m_1^{\text s}, m_2^{\text s}, z^n)\nonumber\\
\leq & n(R_1^{\text o} + R_1^{\text g} + R_2^{\text o} + R_2^{\text g}).
\end{align}
Moreover, based on the definition of $Q(m_1^{\text s}, m_2^{\text s}, z^n)$ in (\ref{Q}), we have
\begin{align}\label{HL1L2Zm1m22}
& H(L_1, L_2| m_1^{\text s}, m_2^{\text s}, O (m_1^{\text s}, m_2^{\text s}, z^n) = 0, z^n)\nonumber\\
\leq & \log(Q(m_1^{\text s}, m_2^{\text s}, z^n))\nonumber\\
\leq & n (\Delta + \delta_1 (\epsilon)) + 1, ~\forall~ z^n \in {\cal T}_\epsilon^{(n)} (Z),
\end{align}
where the last step holds due to the fact that when $O (m_1^{\text s}, m_2^{\text s}, z^n) = 0$, $Q(m_1^{\text s}, m_2^{\text s}, z^n) \leq 2^{n (\Delta + \delta_1 (\epsilon)) + 1}$.
Based on (\ref{HL1L2Zm1m21}), (\ref{HL1L2Zm1m23}) and (\ref{HL1L2Zm1m22}), $H(L_1, L_2| m_1^{\text s}, m_2^{\text s}, Z^n)$ can be upper-bounded as follows
\begin{align}\label{HL1L2Zm1m2}
& H(L_1, L_2| m_1^{\text s}, m_2^{\text s}, Z^n)\nonumber\\
= & P\left\{ Z^n \!\in\! {\cal T}_\epsilon^{(n)} (Z) \right\} H(L_1, L_2| m_1^{\text s}, m_2^{\text s}, Z^n, Z^n \!\in\! {\cal T}_\epsilon^{(n)} (Z))\nonumber\\
+ & P\left\{ Z^n \!\notin\! {\cal T}_\epsilon^{(n)} (Z) \right\} H(L_1, L_2| m_1^{\text s}, m_2^{\text s}, Z^n, Z^n \!\notin\! {\cal T}_\epsilon^{(n)} (Z))\nonumber\\
\leq & \sum_{z^n \in {\cal T}_\epsilon^{(n)} (Z)} p (z^n) H(L_1, L_2| m_1^{\text s}, m_2^{\text s}, z^n) + n \alpha_1\nonumber\\
= & \!\!\sum_{z^n \in {\cal T}_\epsilon^{(n)} (Z)}\!\!\!\!\! \left\{  p_1 (z^n) H(L_1, L_2| m_1^{\text s}, m_2^{\text s}, O (m_1^{\text s}, m_2^{\text s}, z^n) \!=\! 1, z^n) \right.\nonumber\\
+ & \left.  p_2 (z^n) H(L_1, L_2| m_1^{\text s}, m_2^{\text s}, O (m_1^{\text s}, m_2^{\text s}, z^n) \!=\! 0, z^n)\right\} + n \alpha_1\nonumber\\
\leq & \sum_{z^n \in {\cal T}_\epsilon^{(n)} (Z)} \big \{ p (z^n) \alpha_2 H(L_1, L_2| m_1^{\text s}, m_2^{\text s}, z^n) \nonumber\\
+ & p (z^n) H(L_1, L_2| m_1^{\text s}, m_2^{\text s}, O (m_1^{\text s}, m_2^{\text s}, z^n) = 0, z^n) \big \} + n \alpha_1\nonumber\\
\leq & n (\Delta + \delta_2 (\epsilon)),
\end{align}
where
\begin{align}\label{del}
& \alpha_1 = P\left\{ Z^n \notin {\cal T}_\epsilon^{(n)} (Z) \right\} (R_1^{\text o} + R_1^{\text g} + R_2^{\text o} + R_2^{\text g}),\nonumber\\
& p_1 (z^n) = p (z^n) P \left\{ O (m_1^{\text s}, m_2^{\text s}, z^n) = 1 \right\}, ~\forall~ z^n \in {\cal T}_\epsilon^{(n)} (Z)\nonumber\\
& p_2 (z^n) = p (z^n) P \left\{ O (m_1^{\text s}, m_2^{\text s}, z^n) = 0 \right\}, ~\forall~ z^n \in {\cal T}_\epsilon^{(n)} (Z)\nonumber\\
& \alpha_2 = \max \left\{ P \left\{ O (m_1^{\text s}, m_2^{\text s}, z^n) = 1 \right\}, ~\forall~ z^n \in {\cal T}_\epsilon^{(n)} (Z) \right\},\nonumber\\
& \delta_2 (\epsilon) = \delta_1 (\epsilon) + \alpha_2 (R_1^{\text o} + R_1^{\text g} + R_2^{\text o} + R_2^{\text g}) + \frac{1}{n} + \alpha_1.
\end{align}
By the LLN, $P\left\{ Z^n \notin {\cal T}_\epsilon^{(n)} (Z) \right\} \rightarrow 0$ as $n \rightarrow \infty$. 
Hence, $\alpha_1 \rightarrow 0$ as $n \rightarrow \infty$.
In addition, since $P \left\{ O (m_1^{\text s}, m_2^{\text s}, z^n) = 1 \right\} \rightarrow 0, \forall z^n \in {\cal T}_\epsilon^{(n)} (Z)$ as $n \rightarrow \infty$, $\alpha_2 \rightarrow 0$ as $n \rightarrow \infty$. 
$\delta_2 (\epsilon)$ can thus be arbitrarily small as $n \rightarrow \infty$.
Hence,
\begin{align}\label{upper_bound1}
& \lim_{n \rightarrow \infty} \frac{1}{n} H(L_1, L_2| M_1^{\text s}, M_2^{\text s}, Z^n)\nonumber\\
= & \lim_{n \rightarrow \infty} \sum_{m_1^{\text s} = 1}^{2^{n R_1^{\text s}}} \sum_{m_2^{\text s} = 1}^{2^{n R_2^{\text s}}} \frac{1}{n} 2^{-n (R_1^{\text s} + R_2^{\text s})} H(L_1, L_2| m_1^{\text s}, m_2^{\text s}, Z^n)\nonumber\\
\leq & \Delta + \delta.
\end{align}
Theorem \ref{lemma2} is thus proven.

\section{Proof of Theorem \ref{theorem3}}
\label{Appendix_C}

\setcounter{TempEqCnt}{\value{equation}}
\setcounter{equation}{79}
\begin{figure*}[ht]
	\begin{align}\label{N_square}
	& {\mathbb E} \left[(Q(m_1^{\text s}, m_2^{\text s}, z^n))^2\right] = {\mathbb E} \left[\left(\sum_{(l_1,l_2) \in {\cal L}_{1,m_1^{\text s}} \times {\cal L}_{2,m_2^{\text s}} } O' (l_1, l_2)\right)^2\right]\nonumber\\
	& = \sum_{(l_1,l_2) \in {\cal L}_{1,m_1^{\text s}} \times {\cal L}_{2,m_2^{\text s}}} \left\{ P \left\{ (x_1^n (l_1), x_2^n (l_2)) \in {\cal T}_\epsilon^{(n)} (X_1, X_2| z^n) \right\} \right.\nonumber\\
	& + \sum_{l_2' \in {\cal L}_{2,m_2^{\text s}}\setminus \{l_2\} } P \left\{ (x_1^n (l_1), x_2^n (l_2)) \in {\cal T}_\epsilon^{(n)} (X_1, X_2| z^n),~ x_2^n (l_2') \in {\cal T}_\epsilon^{(n)} (X_2| x_1^n  (l_1), z^n) \right\}\nonumber\\
	& + \sum_{l_1' \in {\cal L}_{1,m_1^{\text s}}\setminus \{l_1\} } P \left\{ (x_1^n (l_1), x_2^n (l_2)) \in {\cal T}_\epsilon^{(n)} (X_1, X_2| z^n),~ x_1^n (l_1') \in {\cal T}_\epsilon^{(n)} (X_1| x_2^n  (l_2), z^n) \right\}\nonumber\\
	& + \sum_{(l_1',l_2') \in {\cal L}_{1,m_1^{\text s}} \times {\cal L}_{2,m_2^{\text s}} \setminus \left\{(l_1, l_2)\right\} } \left. P \left\{ (x_1^n (l_1), x_2^n (l_2)), (x_1^n (l_1'), x_2^n (l_2')) \in {\cal T}_\epsilon^{(n)} (X_1, X_2| z^n) \right\} \right\}\nonumber\\
	& = 2^{n (R_1^{\text o} + R_1^{\text g} + R_2^{\text o} + R_2^{\text g})} \left\{ p_1 + (2^{n (R_2^{\text o} + R_2^{\text g})} - 1) p_2 + (2^{n (R_1^{\text o} + R_1^{\text g})} - 1) p_3 + (2^{n(R_1^{\text o} + R_1^{\text g} + R_2^{\text o} + R_2^{\text g})} - 1) p_4 \right\}\nonumber\\
	& \leq 2^{n (\Delta + \delta_1 (\epsilon))} + \sum_{k \in \cal K} 2^{n (2 \Delta - \Delta_k + \delta_1 (\epsilon))} + \left\{{\mathbb E} [Q(m_1^{\text s}, m_2^{\text s}, z^n)]\right\}^2.
	\end{align}
	\hrulefill
\end{figure*}

Using the conditional typicality lemma, for sufficiently large $n$, we have
\setcounter{equation}{68}
\begin{align}
& |{\cal T}_\epsilon^{(n)} (X_1, X_2| z^n)| \leq 2^{n(H(X_1, X_2| Z) + \epsilon)}, \label{T_conditional1}\\
& |{\cal T}_\epsilon^{(n)} (X_2| x_1^n, z^n)| \leq 2^{n(H(X_2| X_1, Z) + \epsilon)},\label{T_conditional4}\\
& |{\cal T}_\epsilon^{(n)} (X_1| x_2^n, z^n)| \leq 2^{n(H(X_1| X_2, Z) + \epsilon)}.\label{T_conditional5}
\end{align}
Let 
\begin{equation}\label{p_1}
p_1 = P \left\{ (X_1^n, X_2^n) \in {\cal T}_\epsilon^{(n)} (X_1, X_2| z^n) \right\}.
\end{equation}
Since $X_1^n$ and $X_2^n$ are independent, an upper bound of $p_1$ can be obtained as follows
\begin{align}\label{p_1_up}
p_1 & = \sum_{(x_1^n, x_2^n) \in {\cal T}_\epsilon^{(n)} (X_1, X_2| z^n)} p(x_1^n) p(x_2^n)\nonumber\\
& \leq 2^{n(H(X_1, X_2| Z) + \epsilon)} 2^{-n(H(X_1) - \epsilon)} 2^{-n(H(X_2) - \epsilon)}\nonumber\\
& \leq 2^{-n(I(X_1, X_2; Z) - \delta_1 (\epsilon))},
\end{align}
where $\delta_1 (\epsilon) = 5 \epsilon$.
Furthermore, denote
\begin{align}\label{p2p3p4}
& p_2 \!=\! P \left\{\! (X_1^n, X_2^n) \!\in\! {\cal T}_\epsilon^{(n)} (X_1, X_2| z^n), {\tilde X}_2^n \!\in\! {\cal T}_\epsilon^{(n)} (X_2| x_1^n, z^n) \!\right\}\!, \nonumber \\
& p_3 \!=\! P \left\{\! (X_1^n, X_2^n) \!\in\! {\cal T}_\epsilon^{(n)} (X_1, X_2| z^n), {\tilde X}_1^n \!\in\! {\cal T}_\epsilon^{(n)} (X_1| x_2^n, z^n) \!\right\}\!, \nonumber \\
& p_4 \!=\! P \left\{ (X_1^n, X_2^n), ({\tilde X}_1^n, {\tilde X}_2^n) \!\in\! {\cal T}_\epsilon^{(n)} (X_1, X_2| z^n) \right\} \!=\! p_1^2,
\end{align}
where ${\cal T}_\epsilon^{(n)} (X_2| x_1^n, z^n)$ and ${\cal T}_\epsilon^{(n)} (X_1| x_2^n, z^n)$ are defined in (\ref{T_conditional2}) and (\ref{T_conditional3}), respectively.
Since $X_1^n$, $X_2^n$ and ${\tilde X}_2^n$ are independent, we have
\begin{align}\label{p_2_up}
p_2 & = \sum_{(x_1^n, x_2^n) \in {\cal T}_\epsilon^{(n)} \!(X_1, X_2| z^n)} \!\!\!p(x_1^n) p(x_2^n) \sum_{ {\tilde x}_2^n \in {\cal T}_\epsilon^{(n)} \!(X_2| x_1^n, z^n)} \!\!\!p({\tilde x}_2^n)\nonumber\\
& \leq 2^{-n(I(X_1, X_2; Z) - 3 \epsilon)} 2^{n(H(X_2| X_1, Z) + \epsilon)} 2^{-n(H(X_2) - \epsilon)}\nonumber\\
& = 2^{-n(I(X_1, X_2; Z) + I(X_2; Z| X_1) - \delta_1 (\epsilon))}.
\end{align}
Similarly, $p_3$ can be upper bounded as follows
\begin{equation}\label{p_3_up}
p_3 \leq 2^{-n(I(X_1, X_2; Z) + I(X_1; Z| X_2) - \delta_1 (\epsilon))}.
\end{equation}

By introducing indicator variable
\begin{equation}\label{indicator_E0}
O' (l_1, l_2) \!=\! \left\{\!\!\!
\begin{array}{ll}
1,&  \!\!\!{\text {if}} ~(x_1^n (l_1), x_2^n (l_2)) \!\in\! {\cal T}_\epsilon^{(n)} (X_1, X_2| z^n),\\
0,&  \!\!\!{\text {otherwise}},\\
\end{array} \right.\!\!\!\!
\end{equation} 
where $z^n \in {\cal T}_\epsilon^{(n)} (Z)$ and $(l_1,l_2) \in {\cal L}_{1,m_1^{\text s}} \times {\cal L}_{2,m_2^{\text s}}$, $Q\left(m_1^{\text s}, m_2^{\text s}, z^n\right)$ can be represented as
\begin{equation}\label{N0}
Q(m_1^{\text s}, m_2^{\text s}, z^n) = \sum_{(l_1,l_2) \in {\cal L}_{1,m_1^{\text s}} \times {\cal L}_{2,m_2^{\text s}} } O' (l_1, l_2).
\end{equation}
Then, we have (\ref{N_expectation}) as follows and (\ref{N_square}) on top of the next page
\begin{align}\label{N_expectation}
{\mathbb E} \left[Q(m_1^{\text s}, m_2^{\text s}, z^n)\right] = & \sum_{(l_1,l_2) \in {\cal L}_{1,m_1^{\text s}} \times {\cal L}_{2,m_2^{\text s}} } {\mathbb E} \left[ O' (l_1, l_2) \right]\nonumber\\
= & \sum_{(l_1,l_2) \in {\cal L}_{1,m_1^{\text s}} \times {\cal L}_{2,m_2^{\text s}} } p_1\nonumber\\
= & |{\cal L}_{1,m_1^{\text s}} \times {\cal L}_{2,m_2^{\text s}}| p_1\nonumber\\
= & 2^{n(R_1^{\text o} + R_1^{\text g} + R_2^{\text o} + R_2^{\text g})} p_1\nonumber\\
\leq & 2^{n (\Delta + \delta_1 (\epsilon))}.
\end{align}
According to (\ref{N_expectation}) and (\ref{N_square}),
\setcounter{equation}{80}
\begin{align}\label{VarN1}
&{\text {Var}} \left[ Q(m_1^{\text s}, m_2^{\text s}, z^n) \right] \nonumber\\
= & {\mathbb E} \left[(Q(m_1^{\text s}, m_2^{\text s}, z^n))^2\right] - \left\{{\mathbb E} [Q(m_1^{\text s}, m_2^{\text s}, z^n)]\right\}^2\nonumber\\
\leq & 2^{n (\Delta + \delta_1 (\epsilon))} + \sum_{k \in \cal K} 2^{n (2 \Delta - \Delta_k + \delta_1 (\epsilon))}.
\end{align}
Theorem \ref{theorem3} is thus proven.

\section{Proof of Lemma \ref{lemma01}}
\label{Prove_lemma01}

For convenience, we consider the two-user case and assume that $\sum_{k \in {\cal S}} I(X_k; Y)> I(X_{\cal S}; Z), ~\forall~ \cal S \subseteq \cal K$ as in Appendix~\ref{Prove_theorem1}. 
Lemma~\ref{lemma01} can be similarly proven as Theorem \ref{theorem1}.
In the following, we show that there exists a $\left( 2^{n R_1^{\text s}}, 2^{n R_1^{\text o}}, 2^{n R_2^{\text s}}, 2^{n R_2^{\text o}}, n \right)$ code such that any rate tuple inside region ${\hat {\mathscr R}} (X_1, X_2)$, i.e., any $(R_1^{\text s}, R_1^{\text o}, R_2^{\text s}, R_2^{\text o})$ satisfying
\begin{equation}\label{rate_region10}
\left\{\!\!\!
\begin{array}{ll}
R_k^{\text s} + R_k^{\text o} < I(X_k; Y) - \epsilon, \forall k \in \cal K, \\
\sum\limits_{k \in {\cal S}} R_k^{\text s} < \sum\limits_{k \in {\cal S}} I(X_k; Y) - I(X_{\cal S}; Z) - \epsilon, \forall \cal S \subseteq \cal K,
\end{array} \right.\!\!
\end{equation}
is achievable. This, together with the standard time-sharing over coding strategies, suffices to prove Lemma \ref{lemma01}. 
We start with the following lemma, which can be easily checked by following the Fourier-Motzkin procedure.
\begin{lemma}\label{lemma02}
	For any rate tuple $(R_1^{\text s}, R_1^{\text o}, R_2^{\text s}, R_2^{\text o})$ satisfying (\ref{rate_region10}), there exists a rate pair $(R_1^{\text g}, R_2^{\text g})$ such that
	\begin{equation}\label{rate_region20}
	\left\{\!\!\!
	\begin{array}{ll}
	R_k^{\text g} \geq 0, ~\forall~ k \in \cal K, \\
	R_k^{\text s} + R_k^{\text o} + R_k^{\text g} < I(X_k; Y) - \epsilon, ~\forall~ k \in \cal K, \\
	\sum\limits_{k \in {\cal S}} (R_k^{\text o} + R_k^{\text g}) \geq I(X_{\cal S}; Z), ~\forall~ \cal S \subseteq \cal K.
	\end{array} \right.
	\end{equation}
\end{lemma}

We generate a codebook and encode the information using the coding scheme provided in Appendix \ref{Prove_theorem1}.
Differently, since independent decoding is assumed, the decoder at the legitimate receiver declares that $({\hat m}_k^{\text s}, {\hat m}_k^{\text o}), ~\forall~ k \in \cal K$ is sent if it is the unique message tuple such that $(x_k^n (l_k), y^n) \in {\cal T}_\epsilon^{(n)}(X_k,Y)$, for some $l_k$ such that $x_k^n (l_k) \in {\cal C}_k ({\hat m}_k^{\text s},{\hat m}_k^{\text o})$.

Since $\sum_{k \in {\cal S}} (R_k^{\text o} + R_k^{\text g}) \geq I(X_{\cal S}; Z), ~\forall~ \cal S \subseteq \cal K$, $\lim_{n \rightarrow \infty} R_{{\text E}, {\cal K}} \leq \delta$ can be proven by following exactly the same steps in Appendix~\ref{A-C}.
Hence, we only need to analyze the probability of error.
Since $R_k^{\text s} + R_k^{\text o} + R_k^{\text g} < I(X_k; Y) - \epsilon, \forall k \in \cal K$, it can be proven by using LLN and the packing lemma that the probability of error averaged over the random codebook and encoding tends to zero as $n \rightarrow \infty$.
Hence, for any $\delta > 0$, we have
\begin{align}
& \lim_{n \rightarrow \infty} P \left\{ \left( {\hat M}_k^{\text s}, {\hat M}_k^{\text o} \right) \neq \left( M_k^{\text s}, M_k^{\text o} \right) \right\} \rightarrow 0\nonumber\\
& \leq 1 - (1 - \delta)^{1/2}, ~\forall~ k \in \cal K.
\end{align}
Then, 
\begin{align}
\lim_{n \rightarrow \infty} P_{\text e} & = 1 - \prod_{k \in \cal K} \left[ 1- \lim_{n \rightarrow \infty} P \left\{ \left( {\hat M}_k^{\text s}, {\hat M}_k^{\text o} \right) \neq \left( M_k^{\text s}, M_k^{\text o} \right) \right\} \right] \nonumber\\
& \leq \delta.
\end{align}
Lemma~\ref{lemma01} for the two-user case is thus proven.
For the case with more than two users, Lemma~\ref{lemma01} can be proven by following similar steps.

\section{Proof of Lemma~\ref{max_R_o}}
\label{Prove_max_R_o}

We first consider the case with two users for convenience and Bob jointly decoding messages.
Assume that $\left(R_1^{\text s}, R_1^{\text o}, R_2^{\text s}, R_2^{\text o}\right)$ is a rate tuple in region ${\mathscr R} (X_1, X_2)$ defined by Theorem~\ref{theorem1} and $R_1^{\text s} + R_2^{\text s} = R_{\text {joint}}^{\text s} (X_{\cal K})$, which is assumed to be positive.
From (\ref{rate_region}) or (\ref{R_joint_DM}), it is known that
\begin{align}\label{R_o_DM_max}
R_1^{\text o} + R_2^{\text o} & \leq R_{\text {joint}} (X_{\cal K}) - R_{\text {joint}}^{\text s} (X_{\cal K})\nonumber\\
& = I(X_{\cal K}; Z),
\end{align}
which indicates that the sum rate at which the users can encode their open messages is no larger than $I(X_{\cal K}; Z)$.

Then, we need to prove that $I(X_{\cal K}; Z)$ is an achievable sum open rate.
Since rate tuple $\left(R_1^{\text s}, R_1^{\text o}, R_2^{\text s}, R_2^{\text o}\right)$ is in region ${\mathscr R} (X_1, X_2)$, it can be proven similarly as Lemma~\ref{lemma0} that there exists a rate pair $(R_1^{\text g}, R_2^{\text g})$ such that
\begin{equation}\label{Fourier_Motzkin}
\left\{\!\!\!
\begin{array}{ll}
R_k^{\text g} \geq 0, ~\forall~ k \in \cal K, \\
\sum\limits_{k \in {\cal S}} (R_k^{\text s} \!+\! R_k^{\text o} \!+\! R_k^{\text g}) \!\leq\! I(X_{\cal S}; Y| X_{\bar {\cal S}}), \forall \cal S \subseteq \cal K, \\
\sum\limits_{k \in {\cal S}} (R_k^{\text o} + R_k^{\text g}) \geq I(X_{\cal S}; Z), \forall \cal S \subseteq \cal K.
\end{array} \right.
\end{equation}
It is obvious from (\ref{Fourier_Motzkin}) that if $R_1^{\text o} + R_2^{\text o} < I(X_{\cal K}; Z)$, we can always split partial rate in $R_k^{\text g}$ to $R_k^{\text o}$, and get ${\hat R}_k^{\text g}$ as well as ${\hat R}_k^{\text o}$, such that
\begin{align}\label{R_o_g_hat}
& {\hat R}_k^{\text g} \geq 0, ~\forall~ k \in \cal K, \nonumber\\
& {\hat R}_k^{\text o} + {\hat R}_k^{\text g} = R_k^{\text o} + R_k^{\text g}, ~\forall~ k \in \cal K,
\end{align}
and
\begin{equation}\label{max_R_o2}
{\hat R}_1^{\text o} + {\hat R}_2^{\text o} = I(X_{\cal K}; Z).
\end{equation}
With (\ref{R_o_g_hat}), it can be easily verified that rate tuple $\left(R_1^{\text s}, {\hat R}_1^{\text o}, {\hat R}_1^{\text g}, \right.$ $\left. R_2^{\text s}, {\hat R}_2^{\text o}, {\hat R}_2^{\text g}\right)$ is in the region defined by (\ref{Fourier_Motzkin}). 
Then, according to the characteristics of Fourier-Motzkin elimination \cite[Appendix D]{el2011network}, we know that rate tuple $\left(R_1^{\text s}, {\hat R}_1^{\text o}, R_2^{\text s}, {\hat R}_2^{\text o} \right)$ is in region ${\mathscr R} (X_1, X_2)$.
In addition, due to (\ref{max_R_o2}), $\left(R_1^{\text s}, {\hat R}_1^{\text o}, R_2^{\text s}, {\hat R}_2^{\text o} \right)$ achieves sum open rate $I(X_{\cal K}; Z)$.
Lemma~\ref{max_R_o} for the two-user case with joint-decoding Bob is thus proven.

For the case with more than two users and the case where Bob independently decodes messages, Lemma~\ref{max_R_o} can be proven by following similar steps.

\section{Proof of Theorem~\ref{theorem_GV_joint}}
\label{prove_lemma3}

We prove Theorem~\ref{theorem_GV_joint} by first establishing a coding scheme for the DM MAC-WT channel with input costs (i.e., introduce input costs to the DM MAC-WT channel in Section \ref{section3}) and then applying the discretization procedure.
For convenience, we consider the two-user case, i.e., ${\cal K} = \{1, 2\}$, and the proof can be naturally extended to the case with more users.

Consider the DM MAC-WT channel $\left({\mathbfcal X}_1, {\mathbfcal X}_2,\right.$ $\left. p(\bm y,\bm z| \bm x_1, \bm x_2), {\mathbfcal Y}, {\mathbfcal Z}\right)$~\footnote{Since we consider Gaussian vector channel in Section \ref{GV_MAC-WT}, we use bold font notations in this appendix to denote vectors or matrices and this has no effect on the proof.}.
Let $b_k (\bm x_k)$ be a nonnegative cost function associated with input vector $\bm x_k \in {\mathbfcal X}_k, \forall k \in {\cal K}$.
For any codeword $\bm x_k^n = (\bm x_{k1}, \cdots, \bm x_{kn})$, assume the following average input cost constraint
\begin{equation}\label{input_cost}
\sum_{i=1}^n b_k (\bm x_{ki}) \leq n P_k, ~\forall~ k \in \cal K.
\end{equation}
Besides, assume w.l.o.g. that there exists a zero-cost vector $\bm x_{k0} \in {\mathbfcal X}_k$ such that $b_k (\bm x_{k0}) = 0$.
Then, based on Theorem \ref{theorem1}, we have the following lemma.
\begin{lemma}\label{lemma5}
	Let $(\bm X_1, \bm X_2, \bm Y, \bm Z) \sim p(\bm x_1) p(\bm x_2) p(\bm y,\bm z| \bm x_1, \bm x_2)$.
	If ${\mathbb E} \left[ b_k(\bm X_k) \right] \leq P_k, \forall k \in \cal K$, then, any rate tuple $\left(R_1^{\text s}, R_1^{\text o},\right.$ $\left. R_2^{\text s}, R_2^{\text o}\right)$ satisfying
	\begin{equation}\label{rate_region5}
	\!\left\{\!\!\!
	\begin{array}{ll}
	\sum\limits_{k \in {\cal S}} (R_k^{\text s} + R_k^{\text o}) \leq I(\bm X_{\cal S}; \bm Y| \bm X_{\bar {\cal S}}), \forall \cal S \subseteq \cal K, \\
	\sum\limits_{k \in {\cal S}} R_k^{\text s} \leq \left[I(\bm X_{\cal S}; \bm Y| \bm X_{\bar {\cal S}}) - I(\bm X_{\cal S}; \bm Z) \right]^+\!\!, \forall \cal S \subseteq \cal K,\\
	R_1^{\text s} \!+\! R_2^{\text s} \!+\! R_k^{\text o} \leq \left[I(\bm X_{\cal K}; \bm Y) \!-\! I(\bm X_{\bar k}; \bm Z) \right]^+\!\!, \forall k \in \cal K,
	\end{array} \right.\!\!
	\end{equation}
	is achievable, where the definition of $\bar {\cal S}$ and $\bar k$ is given in (\ref{rate_region}).
\end{lemma}

\itshape \textbf{Proof:}  \upshape
For the probability density function (pdf) $p(\bm x_1) p(\bm x_2)$ that attains ${\mathbb E} \left[ b_k(\bm X_k) \right] \leq P_k/(1+\epsilon), \forall k \in \cal K$, let ${\cal R} (\bm X_1, \bm X_2)$ denote the set of rate tuples satisfying (\ref{rate_region5}).
Similar to the proof of Theorem \ref{theorem1}, we introduce $\epsilon$ and prove that any rate tuple inside ${\cal R} (\bm X_1, \bm X_2)$ is achievable.
For any rate tuple $(R_1^{\text s}, R_1^{\text o}, R_2^{\text s}, R_2^{\text o})$ inside ${\cal R} (\bm X_1, \bm X_2)$, analogous to Lemma \ref{lemma0}, we can get a rate pair $(R_1^{\text g}, R_2^{\text g})$ such that
\begin{equation}\label{rate_region7}
\left\{\!\!\!
\begin{array}{ll}
R_k^{\text g} \geq 0, ~\forall~ k \in \cal K, \\
\sum\limits_{k \in {\cal S}} (R_k^{\text s} \!+\! R_k^{\text o} \!+\! R_k^{\text g}) \!<\! I(\bm X_{\cal S}; \bm Y| \bm X_{\bar {\cal S}}) \!-\! \epsilon, \forall \cal S \subseteq \cal K, \\
\sum\limits_{k \in {\cal S}} (R_k^{\text o} + R_k^{\text g}) \geq I(\bm X_{\cal S}; \bm Z), \forall \cal S \subseteq \cal K.
\end{array} \right.
\end{equation}
As in Appendix \ref{Prove_theorem1}, for each user $k$, we randomly and independently generate $2^{n(R_k^{\text s}+R_k^{\text o}+R_k^{\text g})}$ codewords $\bm x_k^n (l_k)$ according to $\Pi_{i=1}^n p(\bm x_{ki})$, where $l_k \in {\cal L}_k$, and get a codebook ${\cal C}_k$.
Since ${\mathbb E} \left[ b_k(\bm X_k) \right] \leq P_k/(1+\epsilon)$, if $\bm x_k^n \in {T}_\epsilon^{(n)} (\bm X_k)$, by the typical average lemma \cite[Section 2.4]{el2011network}, $\sum_{i=1}^n b_k (\bm x_{ki}) \leq n P_k/(1+\epsilon)$.
Then, we encode and decode the messages as in Appendix \ref{Prove_theorem1}.
To send message pair $(m_k^{\text s}, m_k^{\text o})$, encoder $k$ chooses a codeword with index $l_k$, i.e., $\bm x_k^n (l_k)$, from codebook ${\cal C}_k$.
What is different here is that to guarantee the input cost, user $k$ transmits $\bm x_k^n (l_k)$ if $\bm x_k^n (l_k) \in {T}_\epsilon^{(n)} (\bm X_k)$.
Otherwise, it transmits $(\bm x_{k0}, \cdots, \bm x_{k0})$.

The next step is to analyze the probability of error and the information leakage rate, which can be realized by following exactly the same steps in Appendix \ref{Prove_theorem1}.
It is thus omitted here.
Hence, any rate tuple inside ${\cal R} (\bm X_1, \bm X_2)$ is achievable.
As $\epsilon \rightarrow 0$, due to the continuity, we know that for the pdf $p(\bm x_1) p(\bm x_2)$ that attains ${\mathbb E} \left[ b_k(\bm X_k) \right] \leq P_k, \forall k \in \cal K$, any rate tuple satisfying (\ref{rate_region5}) is achievable.
Lemma \ref{lemma5} is thus proven.
\hfill $\Box$

In Theorem~\ref{theorem_GV_joint}, we consider Gaussian vector input $\bm X_k \sim {\cal CN}(\bm 0, \bm F_k), \forall k \in \cal K$, with power constraint ${\text {tr}}(\bm F_k) \leq P_k$.
The proof of Theorem~\ref{theorem_GV_joint} for the two-user case can then be achieved by using Lemma \ref{lemma5} and the discretization procedure, i.e., quantizing $\bm X_k$ \cite[Section 3.4]{el2011network}.
The achievability for the more general $K$-user case with $K \geq 1$ can be similarly proven.
Hence, any rate tuple $\left(R_1^{\text s}, R_1^{\text o}, \cdots,\right.$ $\left. R_K^{\text s}, R_K^{\text o}\right)$ satisfying
\begin{align}\label{rate_region8}
\sum_{k \in \cal S} R_k^{\text s} + \sum_{k \in {\cal S}\setminus {\cal S}_1} R_k^{\text o} \leq & \left[I(\bm X_{\cal S}; \bm Y| \bm X_{\bar {\cal S}}) - I(\bm X_{{\cal S}_1}; \bm Z) \right]^+, \nonumber\\
& \forall {\cal S} \subseteq {\cal K} ~{\text {and}}~ {{\cal S}_1} \subseteq \cal S,
\end{align}
is achievable.
Since $\bm X_k \sim {\cal CN}(\bm 0, \bm F_k), \forall k \in \cal K$ and they are independent of each other, we have
\begin{align}\label{mutual}
& I(\bm X_{\cal S}; \bm Y| \bm X_{\bar {\cal S}})\nonumber\\
= & h(\bm Y| \bm X_{\bar {\cal S}}) - h(\bm Y| \bm X_{\cal K})\nonumber\\
= & \log \det \left( \sum_{k \in {\cal S}} \frac{1}{\sigma_B^2} \bm H_k \bm F_k \bm H_k^H + \bm I_B \right),\nonumber\\
& I(\bm X_{{\cal S}_1}; \bm Z)\nonumber\\
= & h(\bm Z) - h(\bm Z| \bm X_{{\cal S}_1})\nonumber\\
= & \log\! \det\!\! \left(\! \sum_{k \in {\cal S}_1} \!\!\bm G_k \bm F_k \bm G_k^H \!\!\left(\! \sum_{j \in {\cal K} \setminus {\cal S}_1} \!\!\!\bm G_j \bm F_j \bm G_j^H \!+\! \sigma_E^2 \bm I_E \!\right)^{-1} \!\!\!+\! \bm I_E \right).
\end{align}
Substituting (\ref{mutual}) into (\ref{rate_region8}), we can get (\ref{rate_region3}).
Theorem~\ref{theorem_GV_joint} is thus proven.

\section{}
\label{Appendix_A}

For brevity, we consider the two-user case.
Since reference \cite{tekin2008general} studied Gaussian scalar MAC-WT channel, we set $B=E=T_k=1, \forall k \in \cal K$ here.
Let $X_k \sim {\cal CN}(0, f_k)$ and $f_k \leq P_k, \forall k \in \cal K$.
Then, (\ref{rate_region3}) can be rewritten as follows
\begin{equation}\label{rate_region9}
\!\left\{\!\!\!
\begin{array}{ll}
\sum\limits_{k \in {\cal S}} (R_k^{\text s} + R_k^{\text o}) \leq I(X_{\cal S}; Y| X_{\bar {\cal S}}), \forall \cal S \subseteq \cal K, \\
\sum\limits_{k \in {\cal S}} R_k^{\text s} \leq \left[I(X_{\cal S}; Y| X_{\bar {\cal S}}) - I(X_{\cal S}; Z) \right]^+\!\!, \forall \cal S \subseteq \cal K,\\
R_1^{\text s} + R_2^{\text s} + R_k^{\text o} \leq \left[I(X_{\cal K}; Y) - I(X_{\bar k}; Z) \right]^+\!\!, \forall k \in \cal K,
\end{array} \right.\!\!
\end{equation}
which has the same formulation as (\ref{rate_region}).
By comparing (\ref{rate_region9}) and (\ref{rate_region11}), we notice that they differ in the third inequality of (\ref{rate_region9}) and (\ref{rate_region11}) gives a larger achievable rate region.

When proving \cite[Theorem 1]{tekin2008general}, it is stated in \cite{tekin2008general} that for any rate tuple $\left(R_1^{\text s}, R_1^{\text o}, R_2^{\text s}, R_2^{\text o}\right)$ satisfying (\ref{rate_region11}), there exists $R_k^x$ such that \cite[eq. (26) -- (28)]{tekin2008general} hold.
For convenience, we rewrite \cite[eq. (26) -- (28)]{tekin2008general} as follows
\begin{equation}\label{rate_region12}
\!\left\{\!\!\!
\begin{array}{ll}
\sum\limits_{k \in {\cal S}} (R_k^{\text s} + R_k^{\text o} + R_k^x) \leq I(X_{\cal S}; Y| X_{\bar {\cal S}}), ~\forall~ \cal S \subseteq \cal K, \\
\sum\limits_{k \in {\cal S}} (R_k^{\text o} + R_k^x) \leq I(X_{\cal S}; Z| X_{\bar {\cal S}}), \forall \cal S \subseteq \cal K,\\
\quad \quad \quad \quad \quad \quad \quad \quad \quad \quad \quad  {\text {with equality if }} \cal S = \cal K,\\
\sum\limits_{k \in {\cal S}} R_k^{\text s} \leq \left[I(X_{\cal S}; Y| X_{\bar {\cal S}}) - I(X_{\cal S}; Z)\right]^+, ~\forall~ \cal S \subseteq \cal K.
\end{array} \right.\!\!
\end{equation}
Note that $R_k^x$ is a key point for the achievability proof of \cite[Theorem 1]{tekin2008general}.
Although not mentioned, it is clear by the definition of $R_k^x$ that
\begin{equation}\label{Rxk}
R_k^x \geq 0, ~\forall~ k \in \cal K.
\end{equation}
When $R_k^{\text o}$ is large, to ensure that \cite[eq. (27)]{tekin2008general}, i.e., the second inequation of (\ref{rate_region12}), is satisfied, some open message of user $k$ can be reclassified as secret message (we call this rate splitting in the following).

In order to check whether it is true that for any rate tuple $\left(R_1^{\text s}, R_1^{\text o}, R_2^{\text s}, R_2^{\text o}\right)$ satisfying (\ref{rate_region11}), with rate splitting, there always exists $R_k^x$ such that (\ref{rate_region12}) and (\ref{Rxk}) hold, we eliminate $R_k^x$ in (\ref{rate_region12}) and (\ref{Rxk}) using the Fourier-Motzkin procedure \cite[Appendix D]{el2011network}, and get
\begin{equation}\label{rate_region21}
\left\{\!\!\!
\begin{array}{ll}
\sum\limits_{k \in {\cal S}} (R_k^{\text s} + R_k^{\text o}) \leq I(X_{\cal S}; Y| X_{\bar {\cal S}}), \forall \cal S \subseteq \cal K, \\
\sum\limits_{k \in {\cal S}} R_k^{\text s} \leq \left[I(X_{\cal S}; Y| X_{\bar {\cal S}}) - I(X_{\cal S}; Z) \right]^+\!\!, \forall \cal S \subseteq \cal K,\\
\sum\limits_{k \in {\cal S}} R_k^{\text o} \leq I(X_{\cal S}; Z| X_{\bar {\cal S}}), \forall \cal S \subseteq \cal K.
\end{array} \right.\!\!
\end{equation}
Denote the sets of rate tuples $(R_1^{\text s}, R_1^{\text o}, R_2^{\text s}, R_2^{\text o})$ satisfying (\ref{rate_region9}), (\ref{rate_region11}) and (\ref{rate_region21}) by $\mathscr R_0$, ${\mathscr R}_1$ and ${\mathscr R}_2$, respectively.
Then, if \cite[Theorem 1]{tekin2008general} is true, with rate splitting, all rate tuples in region ${\mathscr R}_1$ should be able to be transformed to rate tuples in region ${\mathscr R}_2$.
However, in the following we show that with rate splitting, ${\mathscr R}_2$ is equivalent to our region $\mathscr R_0$, and there exist rate tuples in region ${\mathscr R}_1$ which can not be transformed to rate tuples in region ${\mathscr R}_2$.

\subsection{Equivalence Between $\mathscr R_0$ and $\mathscr R_2$ with Rate Splitting}
For any given rate tuple $A_2  = (R_1^{\text s}, R_1^{\text o}, R_2^{\text s}, R_2^{\text o})$ in region ${\mathscr R}_2$, it is obvious from the second and the third inequalities in (\ref{rate_region21}) that 
\begin{equation}
\left\{\!\!\!
\begin{array}{ll}
R_1^{\text s} + R_1^{\text o} + R_2^{\text s} \leq \left[I(X_1, X_2; Y) - I(X_2; Z) \right]^+, \\
R_1^{\text s} + R_2^{\text s} + R_2^{\text o} \leq \left[I(X_1, X_2; Y) - I(X_1; Z) \right]^+.
\end{array} \right.\!\!
\end{equation}
Hence, $A_2$ is also in region $\mathscr R_0$.

\begin{figure}
	\centering
	\includegraphics[width=6cm]{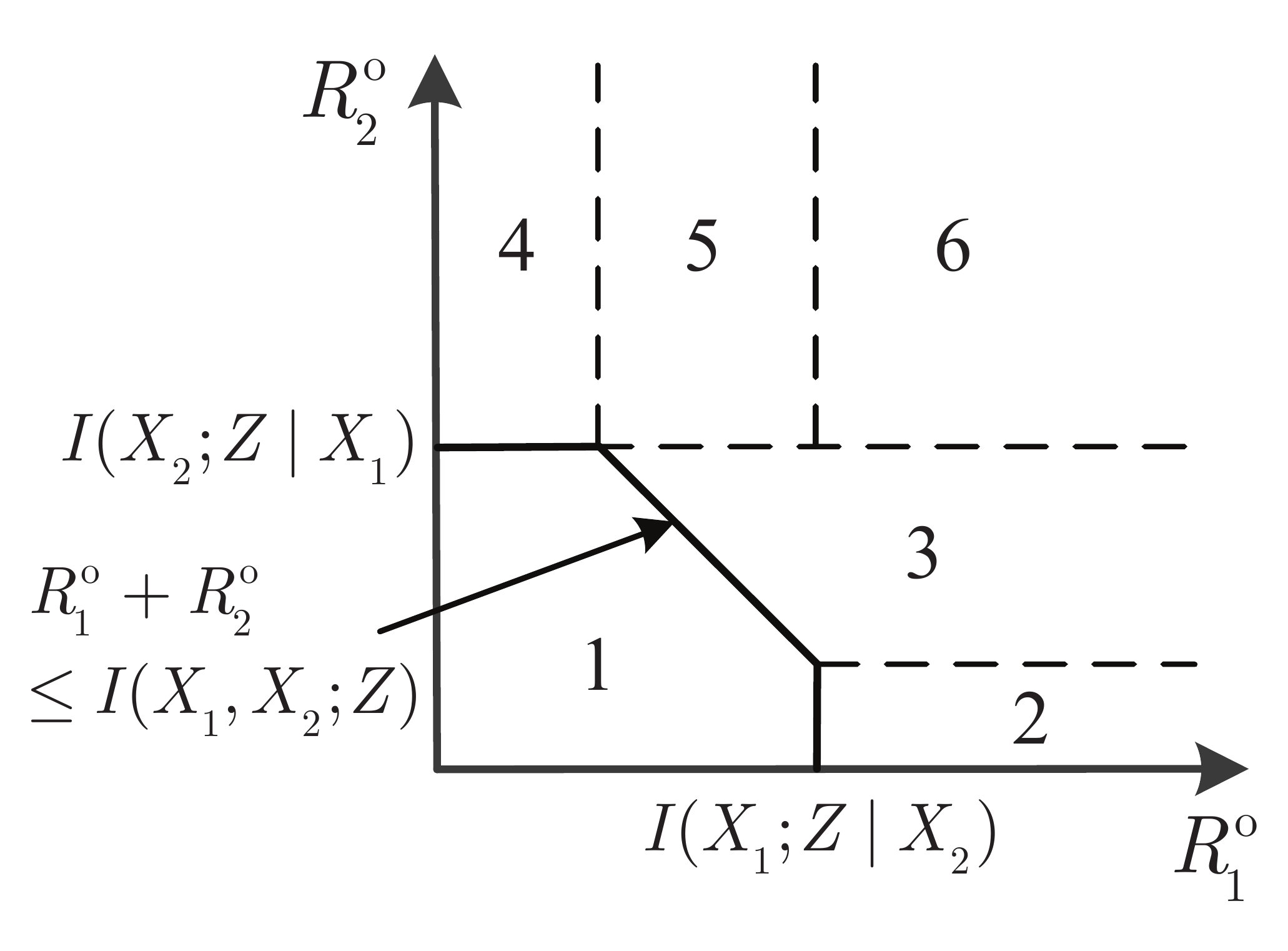}
	\caption{Classification of rate tuples.}
	\label{Fig2}
\end{figure}

Based on the values of $R_1^{\text o}$ and $R_2^{\text o}$, all rate tuples in region $\mathscr R_0$ can be divided into $6$ categories as shown in Fig. \ref{Fig2}. 
In the following we show that, with rate splitting, any given rate tuple $A = (R_1^{\text s}, R_1^{\text o}, R_2^{\text s}, R_2^{\text o})$ in region $\mathscr R_0$ can be transformed to another rate tuple $A'$ in region $\mathscr R_2$.

If rate tuple $A$ belongs to category $1$, i.e.,
\begin{equation}\label{category1}
\sum_{k \in {\cal S}} R_k^{\text o} \leq I(X_{\cal S}; Z| X_{\bar {\cal S}}), \forall \cal S \subseteq \cal K,
\end{equation}
it is obvious from (\ref{rate_region21}) that $A$ is also in region $\mathscr R_2$.

If rate tuple $A$ belongs to category $2$, i.e.,
\begin{equation}\label{category2}
\left\{\!\!\!
\begin{array}{ll}
R_1^{\text o} > I(X_1; Z| X_2), \\
0 \leq R_2^{\text o} \leq I(X_2; Z),
\end{array} \right.
\end{equation}
let
\begin{align}\label{category2_trans}
& {\tilde R}_1^{\text o} = I(X_1; Z| X_2), \nonumber\\
& {\tilde R}_1^{\text s} = R_1^{\text s} + R_1^{\text o} - I(X_1; Z| X_2).
\end{align}
We get a new rate tuple $A' = ({\tilde R}_1^{\text s}, {\tilde R}_1^{\text o}, R_2^{\text s}, R_2^{\text o})$.
Since $A$ is in region $\mathscr R_0$, it satisfies (\ref{rate_region9}).
Hence, 
\begin{align}\label{case2}
& {\tilde R}_1^{\text s} + {\tilde R}_1^{\text o} = R_1^{\text s} + R_1^{\text o}\nonumber\\
& \quad\quad\quad\quad\!\! \leq I(X_1; Y| X_2), \nonumber\\
& {\tilde R}_1^{\text s} + {\tilde R}_1^{\text o} + R_2^{\text s} + R_2^{\text o} = R_1^{\text s} + R_1^{\text o} + R_2^{\text s} + R_2^{\text o}\nonumber\\
& \quad\quad\quad\quad\quad\quad\quad\quad\quad\!\!\! \leq I(X_1, X_2; Y), \nonumber\\
& {\tilde R}_1^{\text s} = R_1^{\text s} + R_1^{\text o} - I(X_1; Y| X_2)\nonumber\\
& \quad \, \leq I(X_1; Y| X_2) - I(X_1; Z| X_2)\nonumber\\
& \quad \, \leq \left[ I(X_1; Y| X_2) - I(X_1; Z) \right]^+,\nonumber\\
& {\tilde R}_1^{\text s} + R_2^{\text s} = R_1^{\text s} + R_1^{\text o} + R_2^{\text s} - I(X_1; Y| X_2) \nonumber\\
& \quad\quad\quad\quad\!\! \leq \left[ I(X_1, X_2; Y) - I(X_2; Z) \right]^+ - I(X_1; Z| X_2)\nonumber\\
& \quad\quad\quad\quad\!\! \leq \left[ I(X_1, X_2; Y) - I(X_1, X_2; Z) \right]^+,\nonumber\\
& {\tilde R}_1^{\text o} + R_2^{\text o} = I(X_1; Z| X_2) + R_2^{\text o}\nonumber\\
& \quad\quad\quad\quad\!\! \leq I(X_1, X_2; Z).
\end{align}
The values of $R_2^{\text s} + R_2^{\text o}$, $R_2^{\text s}$, and $R_2^{\text o}$ remain unchanged.
It is known from (\ref{category2_trans}) and (\ref{case2}) that $A'$ satisfies (\ref{rate_region21}) and is thus in region $\mathscr R_2$.

If $A$ belongs to category $3$, i.e.,
\begin{equation}\label{category3}
\left\{\!\!\!
\begin{array}{ll}
I(X_2; Z) < R_2^{\text o} \leq I(X_2; Z| X_1), \\
R_1^{\text o} + R_2^{\text o} > I(X_1, X_2; Z),
\end{array} \right.
\end{equation}
let
\begin{align}\label{category3_trans}
& {\tilde R}_1^{\text o} = I(X_1, X_2; Z) - R_2^{\text o}, \nonumber\\
& {\tilde R}_1^{\text s} = R_1^{\text s} + R_1^{\text o} + R_2^{\text o} - I(X_1, X_2; Z).
\end{align}
A new rate tuple $A' = ({\tilde R}_1^{\text s}, {\tilde R}_1^{\text o}, R_2^{\text s}, R_2^{\text o})$ is then obtained.
From (\ref{rate_region9}), (\ref{category3}), and (\ref{category3_trans}), we have
\begin{align}
& {\tilde R}_1^{\text s} + {\tilde R}_1^{\text o} = R_1^{\text s} + R_1^{\text o}\nonumber\\
& \quad\quad\quad\quad\!\! \leq I(X_1; Y| X_2), \nonumber\\
& {\tilde R}_1^{\text s} + {\tilde R}_1^{\text o} + R_2^{\text s} + R_2^{\text o} = R_1^{\text s} + R_1^{\text o} + R_2^{\text s} + R_2^{\text o}\nonumber\\
& \quad\quad\quad\quad\quad\quad\quad\quad\quad\!\!\! \leq I(X_1, X_2; Y), \nonumber\\
& {\tilde R}_1^{\text s} = R_1^{\text s} + R_1^{\text o} + R_2^{\text o} - I(X_1, X_2; Z)\nonumber\\
& \quad\, \leq I(X_1; Y| X_2) + R_2^{\text o} - I(X_1; Z) - I(X_2; Z| X_1)\nonumber\\
& \quad\, \leq \left[ I(X_1; Y| X_2) - I(X_1; Z) \right]^+,\nonumber\\
& {\tilde R}_1^{\text s} + R_2^{\text s} = R_1^{\text s} + R_1^{\text o} + R_2^{\text s} + R_2^{\text o} - I(X_1, X_2; Z) \nonumber\\
& \quad\quad\quad\quad\!\! \leq \left[ I(X_1, X_2; Y) - I(X_1, X_2; Z) \right]^+,\nonumber\\
& {\tilde R}_1^{\text o} = I(X_1, X_2; Z) - R_2^{\text o} \nonumber\\
& \quad\, < I(X_1, X_2; Z) - I(X_2; Z) \nonumber\\
& \quad\, = I(X_1; Z| X_2),\nonumber\\
& {\tilde R}_1^{\text o} + R_2^{\text o} = I(X_1, X_2; Z).
\end{align}
The values of $R_2^{\text s} + R_2^{\text o}$, $R_2^{\text s}$, and $R_2^{\text o}$ remain unchanged.
$A'$ thus satisfies (\ref{rate_region21}) and is in region $\mathscr R_2$.

Analogously, if $A$ belongs to category $4$, i.e.,
\begin{equation}\label{category4}
\left\{\!\!\!
\begin{array}{ll}
0 \leq R_1^{\text o} \leq I(X_1; Z), \\
R_2^{\text o} > I(X_2; Z| X_1),
\end{array} \right.
\end{equation}
let
\begin{align}\label{category4_trans}
& {\tilde R}_2^{\text o} = I(X_2; Z| X_1), \nonumber\\
& {\tilde R}_2^{\text s} = R_2^{\text s} + R_2^{\text o} - I(X_2; Z| X_1),
\end{align}
and if $A$ belongs to category $5$, i.e.,
\begin{equation}\label{category5}
\left\{\!\!\!
\begin{array}{ll}
I(X_1; Z) < R_1^{\text o} \leq I(X_1; Z| X_2), \\
R_2^{\text o} > I(X_2; Z| X_1),
\end{array} \right.
\end{equation}
let
\begin{align}\label{category5_trans}
& {\tilde R}_2^{\text o} = I(X_1, X_2; Z) - R_1^{\text o}, \nonumber\\
& {\tilde R}_2^{\text s} = R_1^{\text o} + R_2^{\text s} + R_2^{\text o} - I(X_1, X_2; Z).
\end{align}
It can be similarly proven that the newly obtained rate tuple $A' = (R_1^{\text s}, R_1^{\text o}, {\tilde R}_2^{\text s}, {\tilde R}_2^{\text o})$ is in region $\mathscr R_2$.

If $A$ belongs to category $6$, i.e.,
\begin{equation}\label{category6}
\left\{\!\!\!
\begin{array}{ll}
R_1^{\text o} > I(X_1; Z| X_2), \\
R_2^{\text o} > I(X_2; Z| X_1),
\end{array} \right.
\end{equation}
let
\begin{align}\label{category6_trans}
& {\tilde R}_1^{\text o} = I(X_1; Z| X_2), \nonumber\\
& {\tilde R}_1^{\text s} = R_1^{\text s} + R_1^{\text o} - I(X_1; Z| X_2),\nonumber\\
& {\tilde R}_2^{\text o} = I(X_2; Z), \nonumber\\
& {\tilde R}_2^{\text s} = R_2^{\text s} + R_2^{\text o} - I(X_2; Z).
\end{align}
Then, 
\begin{align}\label{category6_trans1}
& \sum_{k \in {\cal S}} ({\tilde R}_k^{\text s} + {\tilde R}_k^{\text o}) =\sum_{k \in {\cal S}} (R_k^{\text s} + R_k^{\text o})\nonumber\\
& \quad\quad\quad\quad\quad\quad\, \leq I(X_{\cal S}; Y| X_{\bar {\cal S}}), \forall \cal S \subseteq \cal K,\nonumber\\
& {\tilde R}_1^{\text s} \leq I(X_1; Y| X_2) - I(X_1; Z| X_2)\nonumber\\
& \quad\, \leq \left[ I(X_1; Y| X_2) - I(X_1; Z) \right]^+, \nonumber\\
& {\tilde R}_2^{\text s} \leq \left[ I(X_2; Y| X_1) - I(X_2; Z) \right]^+, \nonumber\\
& {\tilde R}_1^{\text s} + {\tilde R}_2^{\text s} = R_1^{\text s} + R_1^{\text o} + R_2^{\text s} + R_2^{\text o} - I(X_1, X_2; Z)\nonumber\\
& \quad\quad\quad\quad \!\!\leq \left[ I(X_1, X_2; Y) - I(X_1, X_2; Z) \right]^+, \nonumber\\
& {\tilde R}_1^{\text o} + {\tilde R}_2^{\text o} = I(X_1, X_2; Z).
\end{align}
Rate tuple $A' =  ({\tilde R}_1^{\text s}, {\tilde R}_1^{\text o}, {\tilde R}_2^{\text s}, {\tilde R}_2^{\text o})$ is thus in region $\mathscr R_2$.

Until now, we have shown that any rate tuple $A_2$ in region $\mathscr R_2$ is also in region $\mathscr R_0$, and by using rate splitting, any rate tuple $A$ in region $\mathscr R_0$ can be transformed to another rate tuple $A'$ in region $\mathscr R_2$.
Therefore, with rate splitting, ${\mathscr R}_2$ is equivalent to $\mathscr R_0$.

\subsection{Unachievable Rate Tuples in ${\mathscr R}_1$}

Next, we show that there exist rate tuples in region ${\mathscr R}_1$ which can not be transformed to rate tuples in region ${\mathscr R}_2$.
Consider rate tuple $A_1 = (R_1^{\text s}, R_1^{\text o}, R_2^{\text s}, R_2^{\text o})$, where
\begin{align}\label{Counterexample}
& R_1^{\text s} = \left[I(X_1; Y| X_2) - I(X_1; Z)\right]^+, \nonumber\\
& R_1^{\text o} = 0,\nonumber\\
& R_2^{\text s} = \left[I(X_2; Y) - I(X_2; Z| X_1)\right]^+, \nonumber\\
& R_2^{\text o} = I(X_1, X_2; Z).
\end{align}
Since $(X_1, X_2, Y, Z) \!\sim~\!p(x_1) p(x_2) p(y,z| x_1, x_2)$, it is possible that 
\begin{align}
& I(X_2; Y) + I(X_1; Z) \leq I(X_2; Y| X_1),\nonumber\\
& I(X_1, X_2; Z) > I(X_2; Z| X_1).
\end{align}
When the above inequality holds, it can be easily found that rate tuple $A_1$ satisfies (\ref{rate_region11}) and is thus in region ${\mathscr R}_1$.
However, since $R_2^{\text o} > I(X_2; Z| X_1)$, (\ref{rate_region21}) is not satisfied, and $A_1$ is thus outside region ${\mathscr R}_2$.
Because $R_1^{\text s} + R_2^{\text s} = \left[I(X_1, X_2; Y) - I(X_1, X_2; Z)\right]^+$, it would be impossible to reduce $R_2^{\text o}$ by increasing $R_2^{\text s}$, i.e., reclassifying some open message of user $2$ as secret message of user $2$, since otherwise $R_1^{\text s} + R_2^{\text s} > \left[I(X_1, X_2; Y) - I(X_1, X_2; Z)\right]^+$.
In this case, rate tuple $A_1$ can not be transformed to another rate tuple in region ${\mathscr R}_1$, indicating that not all rate tuples in region ${\mathscr R}_1$ can be transformed to rate tuples in region ${\mathscr R}_2$ even with rate splitting.

Based on the above analysis, it can be concluded that the statement of \cite[Theorem 1]{tekin2008general} does not hold in general.

\bibliographystyle{IEEEtran}
\bibliography{IEEEabrv,Ref}

\end{document}